\documentclass[10pt,doublecolumn,twoside]{IEEEtran}

\usepackage{etoolbox}
\newtoggle{doublecolumn}
\newtoggle{calculateFigures}

\toggletrue{doublecolumn}
\togglefalse{calculateFigures}
\usepackage{etex}

\usepackage[english]{babel}
\usepackage{lipsum}
\usepackage{amsbsy,amsmath,amsfonts,amssymb,amsthm}
\usepackage{mathtools}
\usepackage{textcomp} 
\usepackage{relsize}

\usepackage{bm,cite,cases,pstricks,pstricks,times,url,verbatim} 
\usepackage[noend]{algpseudocode}
\usepackage{booktabs}
\usepackage{tabularx,array,dcolumn,multirow}

\usepackage{dutchcal} 

\usepackage{soul} 

\usepackage{blkarray,bigdelim} 

\allowdisplaybreaks[4]

\usepackage{centernot} 

\newcommand{\subparagraph}{}

\iftoggle{doublecolumn}{
    \newtheorem{thm}{Theorem}
    \newtheorem{fact}{Fact}
    \newtheorem{lemma}{Lemma}
    \newtheorem{definition}{Definition}
    \newtheorem{conj}{Conjecture}
    \newtheorem{propos}{Proposition}
    \newtheorem{corol}{Corollary}
    \newtheorem{ass}{Assumption}
    \newtheorem{example}{Example}
    \newtheorem{remark}{Remark}
    \newtheorem{note}{Note}
    \newtheorem{obs}{Observation}
}{

    \newtheoremstyle{exampstyle}
      {0} 
      {0} 
      {\itshape} 
      {} 
      {\bfseries} 
      {.} 
      {.5em} 
      {} 

    \theoremstyle{exampstyle} \newtheorem{thm}{Theorem}
    \theoremstyle{exampstyle} 
    \theoremstyle{exampstyle} \newtheorem{lemma}{Lemma}
    \theoremstyle{exampstyle} 
    \theoremstyle{exampstyle} 
    \theoremstyle{exampstyle} \newtheorem{propos}{Proposition}
    \theoremstyle{exampstyle} \newtheorem{corol}{Corollary}
    \theoremstyle{exampstyle} 
    \theoremstyle{exampstyle} \newtheorem{example}{Example}
    \theoremstyle{exampstyle} \newtheorem{remark}{Remark}
    \theoremstyle{exampstyle} 
    \theoremstyle{exampstyle} 
}

\newcommand{\argmax}[1]{\underset{#1}{\operatorname{arg}\,\operatorname{max}}\;}

\makeatletter
\newcommand{\pushright}[1]{\ifmeasuring@#1\else\omit\hfill$\displaystyle#1$\fi\ignorespaces}
\newcommand{\pushleft}[1]{\ifmeasuring@#1\else\omit$\displaystyle#1$\hfill\fi\ignorespaces}

\begingroup
\catcode`\#=11
\gdef\noautorotate{-dAutoRotatePages#/None}
\endgroup

\usepackage{graphicx,xcolor,float,dblfloatfix}
\usepackage{psfrag}

\usepackage{caption}
\usepackage{subcaption}

\newcommand{\subalign}[1]{%
  \vcenter{%
    \Let@ \restore@math@cr \default@tag
    \baselineskip\fontdimen10 \scriptfont\tw@
    \advance\baselineskip\fontdimen12 \scriptfont\tw@
    \lineskip\thr@@\fontdimen8 \scriptfont\thr@@
    \lineskiplimit\lineskip
    \ialign{\hfil$\m@th\textstyle##$&$\m@th\textstyle{}##$\crcr
      #1\crcr
    }%
  }
}

\iftoggle{calculateFigures}{
    \usepackage[crop=off,runs=2,pspdf=\noautorotate]{auto-pst-pdf}
}{
    \usepackage[off]{auto-pst-pdf}
}

\usepackage{hyperref}

\graphicspath{%
{./Figures/}
}

\usepackage{caption}
\begin{document}

\author{\IEEEauthorblockN{Alessandro~Biason,~\IEEEmembership{Student~Member,~IEEE,} and~Michele~Zorzi,~\IEEEmembership{Fellow,~IEEE}
\thanks{The authors are with the Department of Information Engineering, University of
Padova, via Gradenigo 6b, 35131, Padova, Italy. email: \{biasonal,zorzi\}@dei.unipd.it.}
\thanks{A preliminary version of this paper has been presented at European Wireless 2015~\cite{Biason2015a}.}
}}

\title{Joint Transmission and Energy Transfer Policies for Energy Harvesting Devices with Finite Batteries}

\maketitle

\thispagestyle{empty}
\pagestyle{empty}

\begin{abstract}
One of the main concerns in traditional Wireless Sensor Networks (WSNs) is energy efficiency. In
this work, we analyze two techniques that can extend network lifetime. The first is
Ambient \emph{Energy Harvesting} (EH), \emph{i.e.}, the capability of the devices to gather energy
from the environment, whereas the second is Wireless \emph{Energy Transfer} (ET), that can
be used to exchange energy among devices. We study the combination of these techniques, showing
that they can be used jointly to improve the system performance. We consider a transmitter-receiver pair, showing how the ET improvement depends upon the statistics of the energy arrivals and the energy consumption of the devices. With the aim of maximizing a reward function, \emph{e.g.}, the average transmission rate, we find performance upper bounds with and without ET, define both online and offline optimization problems, and present results based on realistic energy arrivals in indoor and outdoor environments. We show that ET can significantly improve the system performance even when a sizable fraction of the transmitted energy is wasted and that, in some scenarios, the online approach can obtain close to optimal performance.
\end{abstract}


\section{Introduction}

\IEEEPARstart{I}{n} the past several years a lot of research has focused on Wireless Sensor Networks (WSNs), where one of the most important questions is how to prolong the network lifetime. In this work we discuss the combination of two different techniques: Ambient \emph{Energy Harvesting} (EH), that allows a device to refill its battery gathering energy from the environment, and Wireless \emph{Energy Transfer} (ET), that makes it possible to exchange energy among different devices. In this paper, we show how ET and EH can be jointly used to improve the overall system performance and prolong network lifetime. Indeed, in some scenarios, a node may receive much more energy and/or consume less energy than some of its neighbors. In these cases, it is reasonable to transmit energy from the rich energy source to other nodes in order to balance the energy levels. ET enables this possibility, and combining it with Energy Harvesting is interesting because it allows to better exploit the renewable energy source and avoid energy overflows.
An example of application is the design of energy-aware routing algorithms that exploit the possibility of sharing energy.

As a first step to understand the key tradeoffs before addressing more complex scenarios, in this paper we consider a network composed of two devices (here we focus on a transmitter and a receiver but the model can be readily extended to the case of two transmitters) equipped with Energy Harvesting and Energy Transfer interfaces. We explicitly take into account the effects of \emph{finite batteries} and, differently from most of the related literature, model the devices energy consumption with \emph{generic functions}. We show that, in the cases where the scenario is unbalanced, \emph{i.e.}, a device harvests much more energy than the other, it is possible to use Wireless Energy Transfer to balance the energy levels of the two devices and, as a consequence, to achieve higher rewards even when a significant fraction of the transmitted energy is wasted. We initially find analytical performance upper bounds with and without ET. Then, we investigate both online and offline approaches and compare them. We present two scenarios with realistic irradiation data showing that ET can be used to increase the average transmission rate. We also describe the effects of finite batteries on the system performance.


The works most closely related to our paper, which studies the combination of EH and ET, are~\cite{Gurakan2013,Gurakan2012,Gurakan2014}, where Gurakan~\emph{et al.} introduced the concept of \emph{energy cooperation}, unifying the study of energy harvesting and energy transfer techniques. They considered a system composed of a few nodes and investigated optimal offline communication schemes. However, none of these papers considered the effects of finite batteries. Also~\cite{Tutuncuoglu2013,Tutuncuoglu2013a} studied the combination of ET and EH with infinite batteries and bi-directional energy transfer. In~\cite{Tutuncuoglu2014} the authors also presented the case of two transmitters with finite batteries. Differently from our work, these papers focused on optimal \emph{offline} transmission policies and assumed ideal energy consumption. A model that considers the circuitry cost was recently published in~\cite{Ni2015}, where a transmitter and a receiver powered by the same power source with infinite batteries can exchange energy.

Energy Harvesting techniques for WSNs have been widely investigated~\cite{Ulukus2015}. In~\cite{Roundy2003}, a survey of energy scavenging methods was presented. \cite{Gakkestad2011} studied the network performance when solar cells are used to receive energy, showing how the harvested energy changes as a function of the latitude, time of the day and season. Analytically, \cite{Lei2009} formulated the problem of maximizing the average value of the reported data using a node with a rechargeable battery. Sharma \emph{et al.} studied heuristic delay-minimizing policies and sufficient stability conditions for an EHD with a data queue~\cite{Sharma2010,Sharma2008,Sharma2011}. 

Energy harvesting receivers were analyzed in~\cite{Mahdavi-Doost2014b,Mahdavi-Doost2013a,Yates2015}, with particular focus on the optimization of the sampling strategies.
Also, \cite{Arafa2014}~considered a transmitter-receiver pair with harvesting capabilities, using energy consumption functions similar to those considered in our work (see Section~\ref{sec:system_model}).
The model that we use in this work is also similar to the one proposed in~\cite{Michelusi2014,Biason2014} for the optimization of an energy harvesting system \textit{without} ET.

Several different technologies for Energy Transfer have been considered so far. In the literature, until recently, the main focus was on \emph{RF Energy Transfer}. This paradigm has been studied for several decades (see~\cite{Brown1984} for a brief history of RF energy transmission). In the last years, RF Energy Transfer was also considered in WSNs~\cite{Kaushik2013,Kim2011a}. In most of the literature, the authors assume to have a sink (a typical example is the \emph{Powercaster Transmitter}~\cite{powercast}) that supplies energy to several passive sensor nodes (equipped with a \emph{Powerharvester Receiver}, for example). One of the main problems studied in this area so far has been the combination of energy and information transmission. Indeed, even if it would be theoretically possible to transmit energy and data simultaneously, this is not feasible with current technology~\cite{Grover2010}.
Therefore, two techniques were developed: Time Splitting (TS) and Power Splitting (PS). In the first case the transferred energy and data are sent at different times. In the second case the transmit power is split: part of it is used for data and the rest for energy. Works such
as~\cite{Timotheou2014} or \cite{Park2013a} studied the optimal power splitting for the PS technique. TS was used in~\cite{Krikidis2012}, where transmission policies for a relay in a topology composed of three nodes (source, relay and destination) were studied. \cite{Kim2011a} proposed a medium access control mechanism based on ET that achieves a high degree of fairness among the devices. Recently, \cite{Niyato2014} studied a network composed of one access point that transmits RF energy to several nodes, with the aim to design an admission control mechanism. In~\cite{Khan2015}, the authors studied the interleaving problems related to transmitting and receiving energy simultaneously, introducing a polling-based MAC protocol. \cite{Coarasa2013} studied the case where some devices (energy-rich sources) move through the network and refill the batteries of the sensors with RF radiation. In~ \cite{Naderi2014} and~\cite{Nintanavongsa2013} the authors introduced an RF-MAC protocol, where nodes request energy from some transmitters, and these cooperate by sending RF energy to those nodes.

However, RF Energy Transfer, due to the radiative nature of the mechanism, has a very low energy efficiency~\cite{Hui2014} or requires line-of-sight and complex tracking systems~\cite{Sample2013}. For these reasons, other techniques were introduced, \emph{e.g.}, \emph{inductive coupling}, that operates at distances less than a wavelength. Clearly, even though this mechanism is very efficient, it cannot be used in a WSN because of the very short operating distance~\cite{Sample2011}. Another emerging technology is \emph{Strongly Coupled Magnetic Resonances} (SCMR) Energy Transfer, which is a compromise between inductive coupling and RF Energy Transfer: it can be used in mid-range applications (order of $2-3$ meters) and provides a relatively high efficiency. In~\cite{Kurs2007}, it was shown that it is possible to power a $60\ \mbox{W}$ light bulb at a distance of $2\ \mbox{m}$ with an efficiency of $40\%$. The authors also extended this work in~\cite{Kurs2010}, showing that SCMR Energy Transfer can be used to power several devices at the same time with high efficiency. This is possible because non-radiative wireless energy transfer is used, which relies on \emph{near-field} magnetic coupling of conductive loops~\cite{Hui2014}. In~\cite{Sample2011}, the authors showed that it is possible to achieve the maximum available energy transfer efficiency regardless of the orientation of the device, as long as the receiver is in the working range of the transmitter. The two main problems related to SCMR Energy Transfer are that: 1) it is necessary to use coils of large size (order of $20\ \mbox{cm}$) and 2) the transmission range is limited to only a few meters. For these reasons, it is reasonable to assume that the devices are fixed, \emph{e.g.}, two devices in adjacent rooms of a building.
Even if SCMR ET seems promising, only a few applications can be found in the literature so far. \cite{Shi2011} considered a vehicle that travels inside a WSN, periodically recharging the nodes (one at a time) wirelessly, and showed that through periodic charges the network may ideally remain operational for an unlimited amount of time. The authors extended the study to multiple transmissions in~\cite{Xie2015}, and a similar technique was also discussed in~\cite{Zhao2014}. Some applications can be found in biomedical implants, \emph{e.g.},~\cite{RamRakhyani2011}, and a wireless charger prototype based on SCMR Energy Transfer was proposed in~\cite{Yao2011}. 

Our contributions in this paper can be summarized as follows. For a transmitter/receiver pair, we present performance upper bounds with and without ET when the energy costs are general functions that can include, \emph{e.g.}, the circuitry costs. The optimal online and offline policies are introduced and characterized. In particular, we use the offline case as a benchmark for our online policies in the finite horizon setting. We show that ET can significantly improve the system performance and that, in some scenarios, the online policies are close to optimal. We also consider the effects of finite batteries, showing that, although the reward improvement depends upon the battery size, it is not necessary to have very large batteries to obtain high gains.

The paper is organized as follows. Section~\ref{sec:system_model} defines the system model we analyzed, and Section~\ref{sec:upper_bounds} provides the performance upper bounds. In Sections~\ref{sec:online} and~\ref{sec:offline} we introduce the online and offline policies, respectively. Section~\ref{sec:long_term} presents the numerical evaluation for the online policies. In Section~\ref{sec:real_data} we analyze two practical examples using realistic irradiation data. Finally, Section~\ref{sec:conclusions} concludes the paper.

\section{System Model and Optimization Problems}\label{sec:system_model}

We study \emph{Energy Harvesting Devices} (EHDs) that, in addition to the capability of gathering energy from the environment, are also able to \emph{transmit} and/or \emph{receive energy} via a Wireless \emph{Energy Transfer} (ET) mechanism. To characterize this technique, we will deal with a pair of EHDs (or simply \emph{devices} or \emph{sensors}) where one device is the transmitter (TX) that sends data to a receiver (RC), whereas RC can send energy to TX (we will comment on the extension to bi-directional ET in Section~\ref{sec:bidirectional}).

We assume a slotted-time system, where slot $k$ corresponds to the time interval $[k-1,k)$, with $k$ a positive integer. Both devices are equipped with some interface that can harvest ambient energy, \emph{e.g.}, from solar light, indoor light, or vibrations. We also assume that the EHDs are temporally synchronized.

TX transmits data packets toward RC and, in every slot, has a new data packet to send. In general, modeling the transmitter and receiver energy costs is a difficult task: to perform a transmission, in addition to the transmit power, also the costs of sensing, pre-processing (coding) and compressing the data~\cite{Tutuncuoglu2012} have to be considered. For packet reception, instead, the main contributions are sampling (demodulation, filtering, quantization), processing (decoding) and storage~\cite{Yates2015}. We simplify the energy consumption models as follows. For reliable communications at rate $R$, TX needs to provide an SNR (thus a transmit power) that depends upon $R$. Similarly, also the reception power depends upon $R$ because of sampling and processing. By combining these concepts, it is possible to establish a relationship between the reception power and the transmit power (see~\cite{Arafa2014}). Formally, we describe the energy consumptions with two generic continuous, increasing and concave\footnote{In this paper, the term ``concave'' will be used to designate concave \emph{downward} functions, \emph{e.g.}, functions with non-positive second derivative.} functions $q^{\rm tx}(P)$, $q^{\rm rc}(P)$, where $P$ is the transmit power and $q^{\rm tx}(0) = q^{\rm rc}(0) = 0$ (when a device is in sleep mode, it is assumed to consume negligible energy). The transmit power used in slot $k$, $P_k \in \Xi \triangleq [0,\rho_{\rm max}]$, is decided at the beginning of each slot.

\begin{example} \label{ex:energy_consumption}
For a transmitter, a common model for the energy function is~\cite{Sharma2010,Gurakan2013}~
\begin{align}
q^{\rm tx}(P) = \sigma^{\rm tx} P. \label{eq:q_tx}
\end{align}

For the receiver, instead, a reasonable approximation is to assume that the energy function is proportional to the transmission rate:~
\begin{align}
q^{\rm rc}(P) = \alpha^{\rm rc} \ln(1+\Lambda P). \label{eq:q_rc}
\end{align}

This model is a good approximation when the circuitry costs are negligible.
Note that in the low-SNR regime, we can approximate $q^{\rm rc}(\cdot)$ as $q^{\rm rc}(P) \approx \sigma^{\rm rc} P$.

$\sigma^{\rm tx}$, $\sigma^{\rm rc}$, $\alpha^{\rm rc}$ are proper constants and $\Lambda$ is an SNR scaling factor.

The contributions of the circuitry costs can be included in this model by adding to~\eqref{eq:q_tx} and~\eqref{eq:q_rc} two terms $\zeta^{\rm tx}(P)$ and $\zeta^{\rm rc}(P)$ that, starting from $0$, increase quickly until constant values in order to preserve the continuity and concavity of $q^{\rm tx}(P)$ and $q^{\rm rc}(P)$. Note that, in the general case, our model allows the circuitry costs for TX and RC to be different.
\end{example}

The amount of energy to be sent with the Energy Transfer mechanism, $D_k\geq 0$, is decided in every slot. The energy received in slot $k$ can be exploited only in a later slot.
In our work we focus on uni-directional energy transfer from RC to TX. We will discuss in Section~\ref{sec:bidirectional} how to extend this hypothesis to the bi-directional case. We assume that only a fraction $\beta$ of the transmitted energy is received, where $\beta \in [0,1]$ is the \emph{energy transfer efficiency}. Note that the $40\%$ efficiency claimed in~\cite{Kurs2007} for a distance of $2\ \mbox{m}$ is only referred to the transmission itself. Indeed, the effective wall-to-load efficiency (ratio between the power extracted from the wall power outlet and the received power) was $15\%$ and, for this reason, in this work we will use a transfer efficiency $\beta = 0.15$ as a baseline.

The devices have finite batteries that can store at most $E_{\rm max}^{\rm tx}$ and $E_{\rm max}^{\rm rc}$ Joule of energy. The randomness of the energy arrivals is described through two independent processes $\{B_k^{\rm tx}\}$ and $\{B_k^{\rm rc}\}$ with some statistics, \emph{e.g.}, deterministic, Bernoulli or truncated geometric. The energy arrival processes have means $\bar{b}^{\rm tx}>0$ and $\bar{b}^{\rm rc}>0$ and the energy harvested in a slot can be exploited only in a later slot.

With the introduced quantities, the evolutions of the two batteries can be described as:~
\begin{alignat}{2}
E_{k+1}^{\rm tx} & = \min\{E_k^{\rm tx}-q^{\rm tx}(P_k) + \beta D_k +B_k^{\rm tx},\ && E_{\rm max}^{\rm tx}\}, \label{eq:E_k_tx} \\
E_{k+1}^{\rm rc} & = \min \{E_k^{\rm rc}-q^{\rm rc}(P_k)- D_k  +B_k^{\rm rc},\ && E_{\rm max}^{\rm rc} \}, \label{eq:E_k_rc}
\end{alignat}

\noindent where $E_k^{\rm tx}$, $E_k^{\rm rc}$ are the energy levels in slot $k$. Since we consider slots of fixed length, in this work we refer to power or energy interchangeably.

With finite batteries, \emph{energy outage} (empty battery) and \emph{energy overflow} (new energy arrivals that cannot be stored due to a fully charged battery) degrade the system performance and have to be considered in the design of a transmission policy~\cite{Michelusi2012}. We assume that the \emph{state of the system} $\mathbf{S}_k = (E_k^{\rm tx}, E_k^{\rm rc})$ is known to both devices at the beginning of every slot, thereby obtaining an upper bound to the achievable system performance, and leave the case of imperfect knowledge for future study. 

\subsection{Optimization Problems} \label{sec:opt_problems}

A policy $\mu$ defines which action to perform in every slot $k$, \emph{i.e.}, how much energy should be used to transmit data ($\mathbf{P} \triangleq \{P_1,P_2,\ldots\}$) and how much energy should be transferred ($\mathbf{D} \triangleq \{D_1,D_2,\ldots\}$).\footnote{The specific structure of $\mu$ depends upon the considered scenario and will be discussed in more detail in Sections~\ref{sec:online} and~\ref{sec:offline}.}

In this work we consider as metrics the average unconstrained rewards in $K$ slots and in the long-term, defined as~
\begin{align}
G_{\mu}^K & \triangleq \frac{1}{K}\sum_{k = 1}^K g(P_k), \label{eq:G_H_q_d_fin}\\
G_{\mu} & \triangleq \liminf_{K \rightarrow \infty} G_{\mu}^K, \label{eq:G_H_q_d}
\end{align}

\noindent where $g(x)$ is a non-decreasing and concave function of $x$. As a baseline, we focus on the \emph{average normalized transmission rate}, obtained when $g(x) = \ln(1+\Lambda x)$, where $\Lambda$ is an SNR scaling factor. 

Optimizing the long-term average reward is a typical assumption because sensors are generally expected to operate for long times in the same scenario. However, our model can be adapted to different reward functions. For example, if the discounted long-term reward were considered, then the optimization techniques would remain the same.

We consider the following optimization problem~
\begin{align}
\mu^\star = \argmax{\mu} G_\mu, \label{eq:mu_star}
\end{align}

\noindent subject to appropriate feasibility constraints (\emph{i.e.}, the transmission power and the transferred energy must be non-negative and must not exceed the energy available in the batteries).
Since the optimization variables and the specific constraints depend upon the chosen approach, this problem will be discussed in more detail in Sections~\ref{sec:online} and~\ref{sec:offline}.

\subsection{Optimization Approaches}\label{sec:roadmap}

In the previous sections we set up the system model and an optimization problem. The goal of this paper is to solve such problem. More precisely, we proceed as follows. Initially, we introduce some performance upper bounds, \emph{i.e.}, upper bounds to $G_\mu$. These do not depend upon the optimization technique. Then, we discuss~\eqref{eq:mu_star} with two approaches (Sections~\ref{sec:online} and~\ref{sec:offline}). 

\subsubsection{\underline{Online approach}}
In this case, in every time slot $k$, the policy chooses an action that depends upon the current state of the system $\mathbf{S}_k$ and upon the energy arrival statistics. In the online case, the output of the optimization process is a set of rules (one for every state of the system) that, given $\mathbf{S}_k$, can be applied to choose the action to perform. In order to model the system as a Markov Decision Process, in Section~\ref{sec:online} we approximate the continuous model with a discrete one.

\subsubsection{\underline{Offline approach}}
In this case, the policies are found by exploiting the non-causal knowledge of the energy arrivals (not only the statistics). In the offline case, the output of the optimization process is a pair of sequences $(\mathbf{P},\mathbf{D})$ that define in every slot from $1$ to $K$ which action to use.

The main focus of our work is on online policies which, though performing worse than offline policies in general, have the important advantage of not requiring non-causal knowledge of the energy arrivals, and are therefore practically usable. Offline policies will be used in Section~\ref{sec:real_data} as a benchmark, showing that in some relevant cases the performance loss incurred by the online approach can be quite small.

\section{Upper Bounds} \label{sec:upper_bounds}

In this section we introduce upper bounds to $G_{\mu}$ for the cases with and without ET. This is an interesting problem because the presented upper bounds are closely approached in several cases of interest and provide an easy characterization of the system reward without performing any optimization. 

They are derived in the \textit{infinite} horizon case, but can be simply reformulated in the \textit{finite} horizon case by changing the long-term means $\bar{b}^{\rm tx}$ and $\bar{b}^{\rm rc}$ with the means in $K$ slots.\footnote{In the following, we find upper bounds based on the \emph{means} of the harvesting processes. Thus, even if we do not explicitly take into account the specific random behavior of the energy arrivals, we are still considering the fact that energy is gathered over time, which is a fundamental feature of EH.} In particular, we will generalize the following intuitive results. As an example, consider $q^{\rm i}(P) = P$ (in the following, ${\rm i} \in \{{\rm tx,rc}\}$) and $\bar{b}^{\rm rc} > \bar{b}^{\rm tx}$ (RC harvests more energy than TX). An upper bound to $G_\mu$ without ET is given by $g(\bar{b}^{\rm tx})$ and is achievable if the devices consume in every slot (except possibly for a vanishing fraction of them) an amount equal to the average harvested energy. This can happen if the battery sizes are infinite or if the batteries are finite and the energy arrivals are deterministic.
Moreover, since $\bar{b}^{\rm rc} > \bar{b}^{\rm tx}$, it may be interesting to use ET to improve the performance (we recall that RC can send energy to TX). In this case an upper bound is given by a balanced combination of the transmitter and receiver average energy arrivals: $g((\beta \bar{b}^{\rm rc} + \bar{b}^{\rm tx})/(\beta+1))$. Note that in this last expression both $\bar{b}^{\rm tx}$ and $\bar{b}^{\rm rc}$ contribute to increasing the upper bound. Also, we remark that the transfer efficiency $\beta$ needs to be considered. These considerations are formalized in the general case in the following (note that, unlike in the above example, we do not impose any constraints on $\bar{b}^{\rm tx}$ and $\bar{b}^{\rm rc}$).

\subsection{Upper Bound without ET}

We first focus on the case without ET. We have the following result.

\begin{thm}[Upper Bound without ET]\label{thm:upper_bound_noET}
If there exist two continuous and increasing functions $\Psi^{\rm tx}(P)$, $\Psi^{\rm rc}(P)$ such that~
\begin{enumerate}
\item $0 \leq \Psi^{\rm tx}(P) \leq q^{\rm tx}(P)$ and $0 \leq \Psi^{\rm rc}(P) \leq q^{\rm rc}(P), \ \forall P \in \Xi$, and
\item $g(\Psi^{{\rm tx}^{-1}}(\cdot))$ and $g(\Psi^{{\rm rc}^{-1}}(\cdot))$ are concave functions,
\end{enumerate}

\noindent then an upper bound for the reward is~
\begin{align}
  G_{\rm u.b.}^{\rm noET} = g\left( \min\left\{ \Psi^{{\rm tx}^{-1}}(\bar{b}^{\rm tx}), \Psi^{{\rm rc}^{-1}}(\bar{b}^{\rm rc}) \right\} \right).\label{eq:G_ub_noET}
\end{align}

If only $\Psi^{\rm i}(P)$ exists, ${\rm i} \in \{{\rm tx,rc}\}$, then an upper bound is $G_{\rm u.b.}^{\rm noET} = g\left( \Psi^{{\rm i}^{-1}}(\bar{b}^{\rm i}) \right)$.

If neither $\Psi^{\rm tx}(P)$ nor $\Psi^{\rm rc}(P)$ exists, then the optimal reward is infinite.

\begin{proof}
See Appendix~\ref{app:upper_bound_noET}.
\end{proof}
\end{thm}

Note that, in the previous theorem, we convert a \emph{power consumption} $\bar{b}^{\rm i}$ to a \emph{reward} in two steps. First, we apply the inverse function $\Psi^{{\rm i}^{-1}}$ to convert the power consumption into a \emph{transmission power}. Then, we apply the function $g(\cdot)$ to the transmission power in order to obtain the corresponding reward.

In practice, $\Psi^{\rm i}(\cdot)$ is an optimistic auxiliary energy consumption function that makes it possible to mathematically obtain~\eqref{eq:G_ub_noET}. Intuitively, the closer $\Psi^{\rm i}(\cdot)$ and $q^{\rm i}(\cdot)$, the tighter the upper bound.

\begin{remark}
If $\Xi$ is bounded, \emph{i.e.}, $\rho_{\rm max} < \infty$, then the conditions of Theorem~\ref{thm:upper_bound_noET} only need to be satisfied for a finite range of $P$, and therefore it is always possible to find $\Psi^{\rm i}(\cdot)$.
\end{remark}

Note that a particular case of the previous remark is obtained when the battery sizes are finite. In this case $\rho_{\rm max}$ is bounded by the maximum battery size (in particular $q^{\rm i}(\rho_{\rm max}) \leq E_{\rm max}^{\rm i}$).

As shown in the following corollaries, there exist cases in which the upper bound of Theorem~\ref{thm:upper_bound_noET} can be achieved.

\begin{corol} \label{corol:q_opt_det_noET}
If $q^{\rm tx}(\cdot) = \Psi^{\rm tx}(\cdot)$ and $q^{{\rm tx}^{-1}}(\bar{b}^{\rm tx}) \leq q^{{\rm rc}^{-1}}(\bar{b}^{\rm rc})$ (TX is the bottleneck) then, in the deterministic energy arrivals case,\footnote{Note that, since we consider i.i.d. energy arrivals, \emph{deterministic}  is equivalent to \emph{constant}.} the upper bound~\eqref{eq:G_ub_noET} is achievable with finite batteries $E_{\rm max}^{\rm tx} \geq \bar{b}^{\rm tx}$, $E_{\rm max}^{\rm rc} \geq q^{\rm rc}(q^{{\rm tx}^{-1}}(\bar{b}^{\rm tx}))$. An optimal policy is~
\begin{align}
P_k &= 
\begin{cases}
q^{{\rm tx}^{-1}}(\bar{b}^{\rm tx}), \quad &\mbox{if } \bar{b}^{\rm tx} \leq E_k^{\rm tx} \mbox{ and }  \\
& \quad q^{\rm rc}(q^{{\rm tx}^{-1}}(\bar{b}^{\rm tx})) \leq E_k^{\rm rc}, \\
0, \quad &\mbox{otherwise}.
\end{cases}  \label{eq:q_opt_det_noET}
\end{align}

The same holds if the roles of TX and RC are exchanged.
\begin{proof}
Let $v = q^{{\rm tx}^{-1}}(\bar{b}^{\rm tx})$.

Assume that at the beginning $E_1^{\rm tx} = E_1^{\rm rc} = 0$. The batteries evolution is the following: $E_2^{\rm tx} = \bar{b}^{\rm tx}$, $E_2^{\rm rc} = \bar{b}^{\rm rc}$. Note that $q^{\rm tx}(v) = \bar{b}^{\rm tx}$ by definition and $q^{\rm rc}(v) \leq \bar{b}^{\rm rc}$ by hypothesis. At $k = 3$, we have: $E_3^{\rm tx} = 2\bar{b}^{\rm tx} -q^{\rm tx}(v) = \bar{b}^{\rm tx}$ (transmit with power $v$ and then harvest an amount of energy exactly equal to $\bar{b}^{\rm tx}$) and $E_3^{\rm rc} = 2\bar{b}^{\rm rc} - q^{\rm rc}(v) \geq \bar{b}^{\rm rc}$. Thus, in every slot, excluding an initial transient, TX can transmit data with power $v$ and RC is always able to receive them, thus the reward per slot is  $g(v)$. In the long-term, the upper bound in~\eqref{eq:G_ub_noET} is achieved.
With different initial states the reasoning is the same.

Note that in the previous considerations we implicitly used the hypotheses $E_{\rm max}^{\rm tx} \geq \bar{b}^{\rm tx}$, $E_{\rm max}^{\rm rc} \geq q^{\rm rc}(q^{{\rm tx}^{-1}}(\bar{b}^{\rm tx}))$, that are necessary to obtain the thesis.
\end{proof}
\end{corol}

The policy of Equation~\eqref{eq:q_opt_det_noET}, possibly excluding an initial transient, consumes all the energy that is received in every slot, and thus achieves the upper bound $g(\bar{b}^{\rm tx})$.

When the battery sizes are infinite, Corollary~\ref{corol:q_opt_det_noET} generalizes to any energy arrival process.

\begin{corol}\label{corol:inf_batt_no_ET}
If $q^{\rm tx}(\cdot) = \Psi^{\rm tx}(\cdot)$, $q^{{\rm tx}^{-1}}(\bar{b}^{\rm tx}) \leq q^{{\rm rc}^{-1}}(\bar{b}^{\rm rc})$ (TX is the bottleneck) and the battery sizes are infinite then the upper bound~\eqref{eq:G_ub_noET} is achievable for any statistics of the energy arrivals. The same holds if the roles of TX and RC are exchanged.
\end{corol}

A formal proof of Corollary~\ref{corol:inf_batt_no_ET} is given in~\cite{Ozel2012b} for the special case of a linear energy consumption model in a single EHD, but can be extended to our case. To show that a reward arbitrarily close to the upper bound can be achieved, a \emph{Save-and-Transmit Scheme} was introduced, where the device does not transmit in an initial transient in order to accumulate enough energy to absorb energy fluctuations, so as to avoid energy outage and manage to consume and receive, on average, the same energy.

\subsection{Upper Bound with ET}\label{sec:ub_with_ET}
We now derive similar results for the case where ET is considered. We introduce two new functions $\bar{c}^{\rm tx}(\cdot)$ and $\bar{c}^{\rm rc}(\cdot)$ defined as follows:~
\begin{align}
\bar{c}^{\rm tx}(\xi) =&\ \bar{b}^{\rm tx} + \beta \bar{b}^{\rm rc} (1-\xi), \label{eq:c_tx}\\
\bar{c}^{\rm rc}(\xi) =&\ \bar{b}^{\rm rc} \xi, \label{eq:c_rc}
\end{align}

\noindent where $\xi \in [0,1]$ is a constant that represents the average fraction of the harvested energy that is transferred with ET under a policy $\mu$. $\bar{c}^{\rm i}(\xi)$ represents the average amount of energy that can be exploited at device ${\rm i} \in \{{\rm tx,rc}\}$ to transmit or receive. In particular, RC transfers part of the harvested energy, thus the residual energy that it can exploit is, on average, only a fraction $\xi$ of the harvested one ($\bar{b}^{\rm rc}$). TX, in addition to its own harvested energy ($\bar{b}^{\rm tx}$), receives the energy that RC transferred (scaled by the energy transfer efficiency $\beta$). One of the key results of the paper is stated in the following theorem.

\begin{thm}[Upper Bound with ET]\label{thm:upper_bound_ET}

Under the hypotheses of Theorem~\ref{thm:upper_bound_noET}, when ET is used, an upper bound for $G_{\mu}$ is~
\begin{align}
G_{\rm u.b.}^{\rm ET} &= g\left( \Psi^{{\rm rc}^{-1}}(\bar{c}^{\rm rc}(\xi^\star)) \right), \label{eq:G_ub_ET}
\end{align}

\noindent where
\begin{itemize}
  \item if $\Psi^{{\rm rc}^{-1}}(\bar{c}^{\rm rc}(1)) \leq \Psi^{{\rm tx}^{-1}}(\bar{c}^{\rm tx}(1))$, then $\xi^\star = 1$;
  \item otherwise, $\xi^\star$ is such that $\Psi^{{\rm tx}^{-1}}(\bar{c}^{\rm tx}(\xi^\star)) = \Psi^{{\rm rc}^{-1}}(\bar{c}^{\rm rc}(\xi^\star))$.
\end{itemize}

\begin{proof}
From Theorem~\ref{thm:upper_bound_noET}, an upper bound is given using $\bar{b}^{\rm tx}$ and $\bar{b}^{\rm rc}$ inside the $\min$ operation. When ET is used, the average amounts of incoming energy at TX and RC are $\bar{c}^{\rm tx}(\xi)$ and $\bar{c}^{\rm rc}(\xi)$, respectively. Thus, when $\xi$ is fixed, an upper bound is~
\begin{align}
  G_{\rm u.b.}^{\rm ET}(\xi) = g\left( \min\left\{ \Psi^{{\rm tx}^{-1}}(\bar{c}^{\rm tx}(\xi)), \Psi^{{\rm rc}^{-1}}(\bar{c}^{\rm rc}(\xi)) \right\} \right). \label{eq:G_ub_proof}
\end{align}

\noindent In practice, we replaced $\bar{b}^{\rm tx}$ and $\bar{b}^{\rm rc}$ with $\bar{c}^{\rm tx}(\xi)$ and $\bar{c}^{\rm rc}(\xi)$ because, with ET, the energy that the devices can exploit is described by $\bar{c}^{\rm tx}(\xi)$ and $\bar{c}^{\rm rc}(\xi)$ (see the description of \eqref{eq:c_tx}-\eqref{eq:c_rc}).

Note that $\Psi^{{\rm i}^{-1}}(\cdot)$ is an increasing and continuous function because $\Psi^{\rm i}(\cdot)$ is increasing and continuous. Moreover, $\partial\bar{c}^{\rm tx}(\xi) / \partial \xi < 0$ and $\partial\bar{c}^{\rm rc}(\xi) / \partial \xi > 0$. Thus, the first argument of the minimum in~\eqref{eq:G_ub_proof} is decreasing in $\xi$, whereas the second one is increasing. The minimum of the two is maximized when they are equal, if this is possible, or otherwise for the maximum value of $\xi$, \emph{i.e.}, $\xi^\star = 1$. Note that, since $\Psi^{{\rm tx}^{-1}}(\bar{c}^{\rm tx}(0)) > \Psi^{{\rm rc}^{-1}}(\bar{c}^{\rm rc}(0))=0$, $\xi^\star$ is equal to one if and only if at $\xi = 1$ we have $\Psi^{{\rm rc}^{-1}}(\bar{c}^{\rm rc}(1)) \leq \Psi^{{\rm tx}^{-1}}(\bar{c}^{\rm tx}(1))$, \emph{i.e.}, $\Psi^{{\rm tx}^{-1}}(\bar{c}^{\rm tx}(\xi))$ and $\Psi^{{\rm rc}^{-1}}(\bar{c}^{\rm rc}(\xi))$ do not have an intersection point in $[0,1)$.
\end{proof}
\end{thm}

Corollaries~\ref{corol:q_opt_det_noET} and~\ref{corol:inf_batt_no_ET} can be generalized as follows.

\begin{corol} \label{corol:q_opt_det_ET}
If $q^{\rm tx}(\cdot) = \Psi^{\rm tx}(\cdot)$ and $q^{\rm rc}(\cdot) = \Psi^{\rm rc}(\cdot)$ then, in the deterministic energy arrivals case, the upper bound~\eqref{eq:G_ub_ET} is achievable with finite batteries $E_{\rm max}^{\rm tx} \geq q^{\rm tx}(q^{{\rm rc}^{-1}}(\bar{c}^{\rm rc}(\xi^\star)))$, $E_{\rm max}^{\rm rc} \geq \bar{b}^{\rm rc}$. An optimal policy is~
\begin{align}
P_k &= 
\begin{cases}
q^{{\rm rc}^{-1}}(\bar{c}^{\rm rc}(\xi^\star)), \quad &\mbox{if } \bar{c}^{\rm rc}(\xi^\star) \leq E_k^{\rm rc} \mbox{ and }  \\
& \ \ q^{\rm tx}(q^{{\rm rc}^{-1}}(\bar{c}^{\rm rc}(\xi^\star))) \leq E_k^{\rm tx}, \\
0, \quad &\mbox{otherwise},
\end{cases}  \label{eq:q_opt_det_ET} \\
D_k &= 
\begin{cases}
\bar{b}^{\rm rc}-q^{\rm rc}(P_k), \quad &\mbox{if } E_k^{\rm rc} \geq \bar{b}^{\rm rc}, \\
0, \quad &\mbox{otherwise}.
\end{cases}  \label{eq:d_opt_det_ET}
\end{align}

\begin{proof}
The proof is similar to that of Corollary~\ref{corol:q_opt_det_noET}. Let $v = q^{{\rm rc}^{-1}}(\bar{c}^{\rm rc}(\xi^\star))$. At $k = 2$, $E_2^{\rm tx} = \bar{b}^{\rm tx}$ and $E_2^{\rm rc} = \bar{b}^{\rm rc}$. 

If $\bar{b}^{\rm tx}$ is greater than or equal to $q^{\rm tx}(v)$, then the policy chooses $P_2 = v$ because $\bar{c}^{\rm rc}(\xi^\star) \leq \bar{b}^{\rm rc}$ by definition of $\bar{c}^{\rm rc}(\cdot)$ and $D_2 = \bar{b}^{\rm rc} - q^{\rm rc}(v)$ because $E_2^{\rm rc} \geq \bar{b}^{\rm rc}$. Note that the sum $q^{\rm rc}(P_2) + D_2$ is equal to $\bar{b}^{\rm rc}$, thus, at $k = 3$, $E_3^{\rm rc} = \bar{b}^{\rm rc}$. Instead, for TX, $E_3^{\rm tx} = \bar{b}^{\rm tx} -q^{\rm tx}(v) + \bar{b}^{\rm tx} + \beta (\bar{b}^{\rm rc} - q^{\rm rc}(v)) = \bar{b}^{\rm tx} -q^{\rm tx}(v) + \bar{c}^{\rm tx}(\xi^\star)$. If $\xi^\star < 1$, then $E_3^{\rm tx} = \bar{b}^{\rm tx}$ because $v = q^{{\rm tx}^{-1}}(\bar{c}^{\rm tx}(\xi^\star))$, otherwise $E_3^{\rm tx} \geq \bar{b}^{\rm tx}$ (see Theorem~\ref{thm:upper_bound_ET}).

If instead $\bar{b}^{\rm tx} < q^{\rm tx}(v)$, the policy chooses $P_2 = 0$ and $D_2 = \bar{b}^{\rm rc}$. Note that, if $\xi^\star = 1$, we have $q^{{\rm rc}^{-1}}(\bar{b}^{\rm rc}) \leq q^{{\rm tx}^{-1}}(\bar{b}^{\rm tx})$ and the inequality chain becomes $q^{{\rm rc}^{-1}}(\bar{b}^{\rm rc}) \leq q^{{\rm tx}^{-1}}(\bar{b}^{\rm tx}) < q^{{\rm rc}^{-1}}(\bar{b}^{\rm rc})$, which is not possible. Thus $\xi^\star$ must be less than $1$ and $q^{{\rm tx}^{-1}}(\bar{c}^{\rm tx}(\xi^\star)) = q^{{\rm rc}^{-1}}(\bar{c}^{\rm rc}(\xi^\star))$ implies that $q^{\rm tx}(v) = \bar{c}^{\rm tx}(\xi^\star) > \bar{b}^{\rm tx}$. At $k = 3$, $E_3^{\rm tx} = 2 \bar{b}^{\rm tx} + \beta \bar{b}^{\rm rc} > \bar{c}^{\rm tx}(\xi^\star) = q^{\rm tx}(v)$ and $E_3^{\rm rc} = \bar{b}^{\rm rc}$. For $k\geq3$, TX always has enough energy to transmit with power $v$.

The previous considerations hold if the battery sizes satisfy the hypotheses of the theorem.
Thus, after an initial transient, the devices always have enough energy to transmit and receive with power $v$ and in the long term the upper bound~\eqref{eq:G_ub_ET} is achieved.
\end{proof}
\end{corol}

\begin{corol}\label{corol:inf_batt_ET}
If $q^{\rm tx}(\cdot) = \Psi^{\rm tx}(\cdot)$, $q^{\rm rc}(\cdot) = \Psi^{\rm rc}(\cdot)$ and the battery sizes are infinite, then the upper bound~\eqref{eq:G_ub_noET} is achievable for any statistics of the energy arrivals.
\begin{proof}
    See Corollary~\ref{corol:inf_batt_no_ET}.
\end{proof}
\end{corol}

The following result establishes when it is beneficial to use ET.

\begin{propos}
If $q^{\rm tx}(\cdot) = \Psi^{\rm tx}(\cdot)$ and $q^{\rm rc}(\cdot) = \Psi^{\rm rc}(\cdot)$, ET always improves the upper bound (\emph{i.e.}, $G_{\rm u.b.}^{\rm ET} > G_{\rm u.b.}^{\rm noET}$) if and only if $\xi^\star < 1$.
\begin{proof}
ET improves the performance if $G_{\rm u.b.}^{\rm noET} < G_{\rm u.b.}^{\rm ET} \Leftrightarrow g\left( \min\left\{ q^{{\rm tx}^{-1}}(\bar{b}^{\rm tx}), q^{{\rm rc}^{-1}}(\bar{b}^{\rm rc}) \right\} \right) < g\left( q^{{\rm rc}^{-1}}(\bar{c}^{\rm rc}(\xi^\star)) \right)$. Since $g(\cdot)$ is an increasing function, the previous condition is equivalent to $\min\left\{ q^{{\rm tx}^{-1}}(\bar{b}^{\rm tx}), q^{{\rm rc}^{-1}}(\bar{b}^{\rm rc}) \right\} < q^{{\rm rc}^{-1}}(\bar{c}^{\rm rc}(\xi^\star))$. 

\begin{itemize}
    \item \emph{(if)} $\xi^\star < 1$ means that (see Theorem~\ref{thm:upper_bound_ET}) $q^{{\rm rc}^{-1}}(\bar{b}^{\rm rc}) > q^{{\rm tx}^{-1}}(\bar{b}^{\rm tx})$, thus the condition becomes $q^{{\rm tx}^{-1}}(\bar{b}^{\rm tx}) < q^{{\rm rc}^{-1}}(\bar{c}^{\rm rc}(\xi^\star))$. Thanks to Theorem~\ref{thm:upper_bound_ET} and to \eqref{eq:c_tx}-\eqref{eq:c_rc}, and since $q^{{\rm tx}^{-1}}(\cdot)$ is increasing, if $\xi^\star < 1$, then $q^{{\rm rc}^{-1}}(\bar{c}(\xi^\star)) = q^{{\rm tx}^{-1}}(\bar{b}^{\rm tx}+\beta \bar{b}^{\rm rc}(1-\xi^\star)) > q^{{\rm tx}^{-1}}(\bar{b}^{\rm tx})$;
    \item \emph{(only if)} Assume $\xi^\star = 1$. In this case $q^{{\rm rc}^{-1}}(\bar{b}^{\rm rc}) \leq q^{{\rm tx}^{-1}}(\bar{b}^{\rm tx})$, which implies $G_{\rm u.b.}^{\rm noET} = g(q^{{\rm rc}^{-1}}(\bar{b}^{\rm rc}))$ and $G_{\rm u.b.}^{\rm ET} = g(q^{{\rm rc}^{-1}}(\bar{b}^{\rm rc}))$, thus ET does not improve the performance upper bound.
\end{itemize}
\vspace{-0.5cm}
\end{proof}
\end{propos}

Note that when $q^{\rm tx}(\cdot) = \Psi^{\rm tx}(\cdot)$ and $q^{\rm rc}(\cdot) = \Psi^{\rm rc}(\cdot)$, $\xi^\star < 1$ is equivalent to $q^{{\rm tx}^{-1}}(\bar{b}^{\rm tx}) < q^{{\rm rc}^{-1}}(\bar{b}^{\rm rc})$. Thus, independently of the transfer efficiency $\beta$, if the average amount of energy harvested per slot at RC ($\bar{b}^{\rm rc}$) corresponds to a transmission power ($q^{{\rm rc}^{-1}}(\bar{b}^{\rm rc})$) greater than what is used at TX ($q^{{\rm tx}^{-1}}(\bar{b}^{\rm tx})$), then the use of ET results in an increased upper bound. When $\xi^\star = 1$, ET cannot provide any improvement because RC is the energy bottleneck and therefore is unable to cooperate with TX. Also, note that the previous considerations also apply to the actual performance for the deterministic energy arrival case (in which the upper bounds can be achieved).

According to the above results, we can identify three main reasons why the upper bounds may not be achieved: 1) The functions $q^{\rm i}(\cdot)$ and $\Psi^{\rm i}(\cdot)$ do not coincide. In this case, the only chance to obtain a better upper bound is to redefine $\Psi^{\rm i}(\cdot)$, if possible. In the following examples we show how to derive $\Psi^{\rm i}(\cdot)$ in several cases of interest. 2) The batteries are small (see Corollaries~\ref{corol:inf_batt_no_ET} and~\ref{corol:inf_batt_ET}). As the battery sizes grow, the performance gets closer to the upper bounds. 3) The time horizon is finite. Indeed, the save and transmit scheme of Corollary~\ref{corol:inf_batt_no_ET} can be applied only if an infinite number of slots are available.

\subsection{Examples}
\begin{example} \label{ex:lin}
Consider the low-SNR regime (see Example~\ref{ex:energy_consumption}). In this case the energy consumptions of both the transmitter and the receiver are linear in $P$. The functions $\Psi^{\rm i}(\cdot)$ can then be taken equal to $q^{\rm i}(\cdot)$ and the upper bounds are~
\begin{align*}
  G_{\rm u.b.}^{\rm noET} &= g\left( \min\left\{ \frac{\bar{b}^{\rm tx}}{\sigma^{\rm tx}}, \frac{\bar{b}^{\rm rc}}{\sigma^{\rm rc}} \right\} \right),\quad G_{\rm u.b.}^{\rm ET} = g\left(\frac{\bar{b}^{\rm rc}}{\sigma^{\rm rc}} \xi^\star \right), \\
  \xi^\star &= \min\left\{1, \frac{\sigma^{\rm rc}}{\bar{b}^{\rm rc}} \frac{\beta \bar{b}^{\rm rc} + \bar{b}^{\rm tx}}{\beta \sigma^{\rm rc}+ \sigma^{\rm tx}}  \right\}.
\end{align*}

\noindent $\xi^\star$ is a linear combination of the average energy arrivals and is used to balance $\bar{c}^{\rm tx}$ and $\bar{c}^{\rm rc}$.

\end{example}

\begin{example}

Another interesting case is $q^{\rm tx} = \sigma^{\rm tx} P$, $q^{\rm rc} = \alpha^{\rm rc} \ln(1+\Lambda P)$ (Equation~\eqref{eq:q_rc}) and $g(x) = \ln(1+\Lambda P)$. Note that $g(\cdot)$ and $q^{\rm rc}(\cdot)$ are proportional and $g(q^{{\rm rc}^{-1}}(x)) = x/\alpha^{\rm rc}$ is concave. Also in this example the functions $\Psi^{\rm i}(\cdot)$ can be taken equal to $q^{\rm i}(\cdot)$. The upper bounds become~
\begin{align*}
  G_{\rm u.b.}^{\rm noET} &= \min\left\{g\left(  \frac{\bar{b}^{\rm tx}}{\sigma^{\rm tx}}\right),\frac{\bar{b}^{\rm rc}}{\alpha^{\rm rc}}\right\},\quad   G_{\rm u.b.}^{\rm ET} = \frac{\bar{c}^{\rm rc}(\xi^\star)}{\alpha^{\rm rc}},
\end{align*}

\noindent where $\xi^\star$ is the unique solution of~
\begin{align*}
    \frac{\bar{b}^{\rm tx} + \beta \bar{b}^{\rm rc} (1-\xi)}{\sigma^{\rm tx}} = \frac{e^{\xi \bar{b}^{\rm rc}/\alpha^{\rm rc}}-1}{\Lambda},
\end{align*}

\noindent if $\Lambda\bar{b}^{\rm tx}/\sigma^{\rm tx} < e^{ b^{\rm rc}/\alpha^{\rm rc}}-1$, and $\xi^\star = 1$ otherwise.

\end{example}

\begin{example} \label{ex:ex3}

We now want to show a case where $\Psi^{\rm i}(\cdot)$ and $q^{\rm i}(\cdot)$ are not the same. Consider $g(x) = \ln(1+\Lambda x)$, $\bar{b} \triangleq \bar{b}^{\rm tx} = \bar{b}^{\rm rc}$ and $q(\cdot) \triangleq q^{\rm tx}(\cdot) = q^{\rm rc}(\cdot)$ with~
\begin{align}
q(P) =
\begin{cases}
\frac{\zeta+P_n}{P_n} P, \quad & \mbox{if } P < P_n, \\
\zeta + P, \quad & \mbox{if } P \geq P_n,
\end{cases} \label{eq:q_ex3}
\end{align}

\noindent with $P_n$ arbitrarily close to $0$. Note that this energy consumption model is suitable for the cases where the circuitry costs are not negligible.
If we choose $\Psi(P) = q(P)$, then it can be verified that there exist values of $\zeta$ and $\bar{b}$ such that $g(q^{-1}(\cdot))$ is not concave. In this case $g(q^{-1}(\bar{b}))$ is not guaranteed to be an upper bound.

\begin{figure}
  \centering
  \includegraphics[width=1\columnwidth]{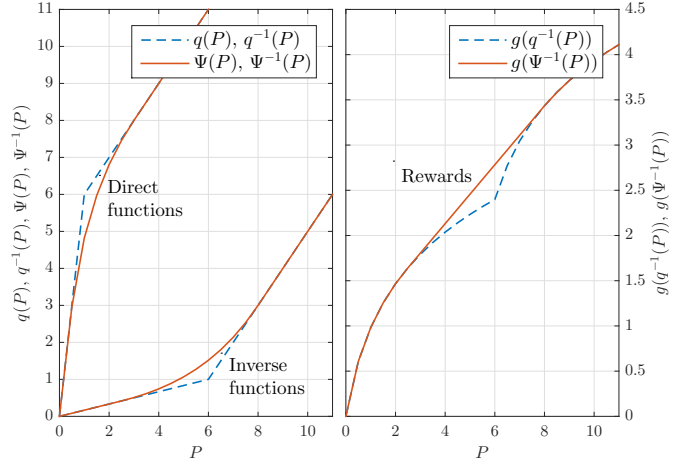}
  \caption{$q(P)$, $\Psi(P)$, their inverse functions, and $g(q^{-1}(P))$, $g(\Psi^{-1}(P))$ of Example~\ref{ex:ex3} as a function of $P$.}
  \label{fig:ex3}
\end{figure}

However, an upper bound can be found by considering a function $\Psi(P)$ defined as in Theorem~\ref{thm:upper_bound_noET}. In \figurename~\ref{fig:ex3} we plot $q(P)$, $\Psi(P)$ and their inverse functions when $\Lambda = 10$, $\zeta = 5$, $E_{\rm max} = 11$. For the purpose of illustration, we arbitrarily set $P_n = 1$. Note that $g(q^{-1}(P))$ is piece-wise concave whereas $g(\Psi^{-1}(P))$ is always concave. The function $\Psi(P)$ is such that $g(\Psi^{-1}(P))$ is divided in three regions. The two external regions are equal to two concave portions of $g(q^{-1}(P))$. The central region is designed to be concave where $g(q^{-1}(P))$ is not, and is obtained considering the straight line that is tangent to  $g(q^{-1}(P))$ in two points without intersecting it. In Section~\ref{sec:long_term} we show that the upper bound given by this choice of $\Psi(P)$ is close to the real performance.
\end{example}

\subsection{Extension: Bi-directional Energy Transfer}\label{sec:bidirectional}

In the following, we present how our model can be extended to the bi-directional ET case. In this context, also TX is able to send part of its stored energy to RC when appropriate. In slot $k$, TX sends an amount of energy $D_k^{{\rm tx}\rightarrow {\rm rc}}$ to RC, whereas RC sends $D_k^{{\rm rc}\rightarrow {\rm tx}}$ to TX. The first term inside the minimum of Equation~\eqref{eq:E_k_tx} has to be changed to~
\begin{align}
    E_k^{\rm tx}-q^{\rm tx}(P_k) + \beta^{{\rm rc}\rightarrow {\rm tx}} D_k^{{\rm rc}\rightarrow {\rm tx}} - D_k^{{\rm tx}\rightarrow {\rm rc}} +B_k^{\rm tx}
\end{align}

\noindent and similarly for Equation~\eqref{eq:E_k_rc} by switching ``tx'' and ``rc''.

The optimization of Equation~\eqref{eq:mu_star} in this case provides three quantities, \emph{i.e.}, the transmission power, the energy sent from RC to TX and vice-versa.\footnote{Note that for any realistic system, in which $\beta^{{\rm rc}\rightarrow {\rm tx}}<1$ and $\beta^{{\rm tx}\rightarrow {\rm rc}} < 1$, under the optimal policy we must have $D_k^{{\rm tx}\rightarrow{\rm rc}}D_k^{{\rm rc}\rightarrow{\rm tx}}=0$, \emph{i.e.}, transferring non-zero energy in both directions simultaneously is strictly sub-optimal.}

The upper bound of Equation~\eqref{eq:G_ub_noET} does not change because it does not depend upon ET. Theorem~\ref{thm:upper_bound_ET} can be reformulated by changing Equations~\eqref{eq:c_tx}-\eqref{eq:c_rc} as follows~
\begin{align}
    \bar{c}^{\rm tx}(\xi^{{\rm tx}\rightarrow {\rm rc}},\xi^{{\rm rc}\rightarrow {\rm tx}}) =&\ \bar{b}^{\rm tx} \xi^{{\rm tx}\rightarrow {\rm rc}} + \beta^{{\rm rc}\rightarrow {\rm tx}} \bar{b}^{\rm rc} (1-\xi^{{\rm rc}\rightarrow {\rm tx}}) ,\\
    \bar{c}^{\rm rc}(\xi^{{\rm tx}\rightarrow {\rm rc}},\xi^{{\rm rc}\rightarrow {\rm tx}}) =&\ \bar{b}^{\rm rc} \xi^{{\rm rc}\rightarrow {\rm tx}} + \beta^{{\rm tx}\rightarrow {\rm rc}} \bar{b}^{\rm tx} (1-\xi^{{\rm tx}\rightarrow {\rm rc}}),
\end{align}

\noindent where $\xi^{{\rm i}\rightarrow {\rm j}}$ represents the average fraction of the harvested energy that is sent from device i to device j.

In our work we decided to focus on the uni-directional case and outline in this section how to extend it for presentation simplicity. Moreover, uni-directional ET can be effectively used in the practically relevant cases where one device harvests more energy than the other. Finally, uni-directional ET can be seen as a simpler lower bound for the bi-directional case.

\section{Online Optimization}\label{sec:online}

We now discuss the online approach and focus on long-term optimization. According to Section~\ref{sec:roadmap}, the aim of an online policy is to define a set of rules that, given the state of the system in a slot, specifies which action (transmission power and transferred energy) should be used in that slot. The online approach is interesting because it requires only a statistical knowledge of the energy arrival process, thus may be effectively used in practice.

In order to formulate the problem as a discrete Markov Decision Process (for which there exist efficient solving algorithms), we introduce the notion of \emph{energy quanta}, \emph{i.e.}, we discretize the amounts of energy (energy arrivals, energy consumptions, energy stored, energy exchanged).\footnote{The accuracy of the discrete approximation of the continuous case can always be improved by using a finer quantization, which however results in a model with more states and therefore higher complexity.} 
The batteries have integer sizes $e_{\rm max}^{\rm tx}$, $e_{\rm max}^{\rm rc}$ and can be considered as buffers.
In order to obtain a consistent formulation, the values of $e_{\rm max}^{\rm tx}$ and $e_{\rm max}^{\rm rc}$ are chosen such that $E_{\rm max}^{\rm tx}/e_{\rm max}^{\rm tx} = E_{\rm max}^{\rm rc}/e_{\rm max}^{\rm rc}$. Under this assumption, one energy quantum corresponds to $E_{\rm max}^{\rm i}/e_{\rm max}^{\rm i}$ Joule. Therefore, when we deal with the online model, all the energy values ($D_k$, $q^{\rm i}(P_k)$, $\bar{b}$, etc.) are expressed as a (not necessarily integer) number of energy quanta.

With the above formulation, we will model the system as a finite two-dimensional Markov Chain (MC). When the MC is in state $\mathbf{e} \triangleq (e^{\rm tx},e^{\rm rc})$, TX and RC have $e^{\rm tx}$ and $e^{\rm rc}$ energy quanta stored in their batteries, respectively. In every state of the MC, a decision is made on the transmission power $\rho(\mathbf{e}) \in \Xi$ of TX and on how many energy quanta RC transfers to TX, namely $d(\mathbf{e}) \in \{0,1,\ldots,e^{\rm rc}_{\rm max}\}$.

Following the approach of~\cite{Biason2015c,Michelusi2012,Tutuncuoglu2015}, in this paper we only consider deterministic policies. Therefore, an online policy $\eta$ specifies a mapping between the current state of the system, $\mathbf{e}$, and the corresponding action (transmitted power $\rho(\mathbf{e})$ and transferred energy $d(\mathbf{e})$), i.e., $\eta = \{(\rho(\mathbf{e}),d(\mathbf{e})),\forall \mathbf{e}\}$. Through an online policy $\eta$, a specific sequence of energy arrivals can be simply mapped to a sequence of actions $(\mathbf{P},\mathbf{D})$.

The batteries evolution~\eqref{eq:E_k_tx}-\eqref{eq:E_k_rc} can be rewritten in terms of energy quanta where, instead of $\beta D_k$ and  $q^{\rm i}(P_k)$, we use $\lfloor\beta D_k\rfloor$ and $q_d^{\rm i}(P_k) \triangleq \lceil q^{\rm i}(\rho(\mathbf{e})) \rceil$, respectively. This choice will result in a lower bound for the real performance (however, we verified that the upper bound obtained by using $\lceil\beta D_k\rceil$ and $\lfloor q^{\rm i}(\rho(\mathbf{e})) \rfloor$ is very similar).

We restrict our study to the set of feasible policies, \emph{i.e.}, those in which, for every $\mathbf{e}$, we have $\rho(\mathbf{e}) \geq 0$, $d(\mathbf{e}) \geq 0$, $q_d^{\rm tx}(\rho(\mathbf{e})) \leq e^{\rm tx}$, $q_d^{\rm rc}(\rho(\mathbf{e})) + d(\mathbf{e}) \leq e^{\rm rc}$.

The reward of Equation~\eqref{eq:G_H_q_d} does not depend upon the starting state when the underlying MC has a unique recurrent class~\cite{Levin2009}. Under this assumption, the long-term reward can be rewritten as~
\begin{align}
  G_{\eta} &= \sum_{e^{\rm tx} = 0}^{e_{\rm max}^{\rm tx}}\sum_{e^{\rm rc} = 0}^{e_{\rm max}^{\rm
  rc}} \pi_{\eta}(\mathbf{e}) g(\rho(\mathbf{e})), \label{eq:G_eta}
\end{align}

\noindent where $\pi_{\eta}(\mathbf{e})$ is the steady-state probability of being in state $\mathbf{e}$ under policy $\eta$. The optimization variables of Problem~\eqref{eq:mu_star} become $(\rho(\mathbf{e}),d(\mathbf{e})),\forall \mathbf{e}$ and the maximization is performed over all the feasible policies.

The Optimal Online Policy $\eta^\star$ (OP-ON) that maximizes $G_{\eta}$ can be found numerically with the Policy Iteration Algorithm (PIA) \cite{Bertsekas2005} by exploiting the full energy arrivals statistics. The algorithm starts with an initial policy (thus we arbitrarily initialize $\rho(\mathbf{e})$ and $d(\mathbf{e})$) and then performs the \emph{policy evaluation} and \emph{improvement} steps in order to iteratively find a new policy, until the reward function $G_{\eta}$ converges (for additional details see~\cite[Section.~7.2]{Bertsekas2005}).

\subsection{Low Complexity Policies} \label{sec:LCP}

In addition to the optimal online policy OP-ON, here we also introduce some simple heuristic policies, that will be used in the numerical evaluations in Sections~\ref{sec:long_term} and~\ref{sec:real_data} to show that, even when sub-optimal policies are adopted, the system reward can be improved using ET. 

In previous works, we studied sub-optimal low complexity policies for EHDs in several cases~\cite{Biason2014},~\cite{Michelusi2012}. However, when EH is combined with ET, the structure of the optimal policy is complex and, moreover, depends upon the energy arrival processes and the energy consumption functions. For these reasons, it is difficult to introduce a simple policy that approximates the optimal one in a broad range of values and so the approaches of~\cite{Biason2014,Michelusi2012} cannot be directly applied.

In the general case, we define the \emph{Greedy Policy} (GP) as follows\footnote{Note that, differently from $q^{\rm i}(\cdot)$, the function $q_d^{\rm i}(\cdot)$ may not be bijective. In this context we define ${q_d^{\rm i}}^{-1}(x) \triangleq \max_{P : q_d^{\rm i}(P) = x} \rho$, \emph{i.e.}, ${q_d^{\rm i}}^{-1}(x)$ is the greatest element of $\Xi$ that is mapped to $x$. This is a reasonable choice because, for all values of $P$ such that $q_d^{\rm i}(P) = x$, the energy consumption is the same but the reward $g(P)$ is different and, since $g(P)$ increases with $P$, we choose the greatest value in order to obtain the highest reward.}~
\begin{align}  
  \begin{split}
    \rho(\mathbf{e}) &= \min\left\{{q_d^{\rm tx}}^{-1}(e^{\rm tx}), {q_d^{\rm rc}}^{-1}(e^{\rm rc})\right\}, \\
    d(\mathbf{e}) &= e^{\rm rc}-q_d^{\rm rc}(\rho(\mathbf{e})).
  \end{split}
  \label{eq:GP}
\end{align}

GP is a simple policy that empties at least one battery in every slot and is independent of the energy arrivals. Consider now the case where both TX and RC have $q^{\rm i}(\cdot) = \Psi^{\rm i}(\cdot)$. We introduce two other policies, namely BP and LCP, as extensions of GP.

The \emph{Balanced Policy} (BP) is defined as the solution of the following system (note that BP does not depend upon the energy arrival statistics, a useful feature when the harvesting process is unknown)~
\begin{align}
\begin{cases}
    E_k^{\rm tx}+ \beta D_k - q^{\rm tx}(P_k) = E_k^{\rm rc} - D_k -q^{\rm rc}(P_k),\\
    P_k = \min\left\{q^{{\rm tx}^{-1}}(E_k^{\rm tx}), q^{{\rm rc}^{-1}}(E_k^{\rm rc}-D_k)\right\}.
\end{cases} \label{eq:BP} 
\end{align}

Instead, the \emph{Low Complexity Policy} (LCP) is defined as follows~
\begin{align}  
  \begin{split}
    \rho(\mathbf{e}) &= \min\left\{{q_d^{\rm tx}}^{-1}(e^{\rm tx}), {q_d^{\rm rc}}^{-1}(e^{\rm rc}), \left[q^{{\rm rc}^{-1}}(\bar{c}^{\rm rc}(\xi^\star))\right]\right\}, \\
    d(\mathbf{e}) &= \min\{e^{\rm rc}-q_d^{\rm rc}(\rho(\mathbf{e})),\left[\bar{b}^{\rm rc}\right]-q_d^{\rm rc}(\rho(\mathbf{e}))\},
  \end{split}
  \label{eq:LCP}
\end{align}

\noindent where $[\cdot] = Round(\cdot)$.

In order to explain how to derive BP according to the above definition, we neglect the \emph{floor} and \emph{ceiling} operations that should be considered in the battery update formulas in the discrete model. At the end of slot $k$, neglecting outage and overflow, the energy levels of the two devices are: $B_k^{\rm tx}+E_k^{\rm tx}+ \beta D_k - q^{\rm tx}(P_k)$ and $B_k^{\rm rc}+E_k^{\rm rc} - D_k -q^{\rm rc}(P_k)$. We impose that at the beginning of the next slot these two quantities be equal. Note that $B_k^{\rm tx}$ and $B_k^{\rm rc}$ are not known a priori,\footnote{It is possible to relax this hypothesis if the arrival process is predictable or partially predictable.} thus we neglect them as well (it is possible to include only the \emph{means} of the energy arrivals, but we verified that this refinement would not provide any significant benefit). Also, since we need to specify both $P_k$ and $D_k$, we need an additional equation. We impose that one of the two batteries is emptied in every slot, and therefore choose $P_k$ as the minimum between $q^{{\rm tx}^{-1}}(E_k^{\rm tx})$ and $q^{{\rm rc}^{-1}}(E_k^{\rm rc}-D_k)$. 

Assume that an acceptable solution of~\eqref{eq:BP} exists and name it $(\bar{\rho},\bar{d})$. Two cases have to be considered:~
\begin{enumerate}
    \item $q^{{\rm tx}^{-1}}(E_k^{\rm tx}) < q^{{\rm rc}^{-1}}(E_k^{\rm rc}-\bar{d}) \Leftrightarrow \bar{\rho} = q^{{\rm tx}^{-1}}(E_k^{\rm tx})$. In this case, the first equation can be simplified and we find $\bar{d} = \frac{E_k^{\rm rc} - q^{\rm rc}(q^{{\rm tx}^{-1}}(E_k^{\rm tx}))}{\beta+1}$;
    \item $q^{{\rm tx}^{-1}}(E_k^{\rm tx}) \geq q^{{\rm rc}^{-1}}(E_k^{\rm rc}-\bar{d}) \Leftrightarrow \bar{\rho} = q^{{\rm rc}^{-1}}(E_k^{\rm rc}-\bar{d})$. In this case $\bar{\rho}$ and $\bar{d}$ can be numerically found.
\end{enumerate}

Also, it may happen that the system does not have acceptable solutions, \emph{i.e.}, $\bar{\rho}$ or $\bar{d}$ is negative or exceeds the current battery levels. In this case we proceed as follows. First, we substitute the second equation into the first one. Then, we find the solution of the first equation, namely $\bar{d}$, following the previous reasoning, \emph{i.e.}, considering the two possible cases. Finally, if $\bar{d}$ is negative, we set $\bar{d} = 0$. Instead, if $\bar{d} > E_k^{\rm rc}$, we set $\bar{d} = E_k^{\rm rc}$. $\bar{\rho}$ is then derived from the second equation.

Once $(\bar{\rho},\bar{d})$ is specified, we extract the online policy as follows (replace $E_k^{\rm i}$ with $e^{\rm i}$): $\rho(\mathbf{e}) = \bar{\rho}$ and $d(\mathbf{e}) = \lfloor \bar{d} \rfloor$. We used the \emph{floor} operation in order to guarantee $q^{\rm rc}(\rho(\mathbf{e})) + d(\mathbf{e}) \leq e^{\rm rc}$ (with the \emph{round} operation, the condition might not be satisfied).

The \emph{Balanced Policy} (BP), obtained according to the above procedure, is designed with the goal to balance the energy levels of the two devices.

The Low Complexity Policy is specified in~\eqref{eq:LCP}. Consider the last terms of the two $\min$ operations. It can be seen that they are the discretized versions of Equations~\eqref{eq:q_opt_det_ET}-\eqref{eq:d_opt_det_ET} (we applied the \emph{round} operations in order to obtain two integer values). Note that the policy in~\eqref{eq:q_opt_det_ET}-\eqref{eq:d_opt_det_ET} does not transmit when the batteries cannot support the use of a power $q^{{\rm rc}^{-1}}(\bar{c}^{\rm rc}(\xi^\star))$, whereas in this case LCP would instead always use the maximum transmit power allowed by the status of the two batteries, which results in the full discharge of at least one of them. Although different from~\eqref{eq:q_opt_det_ET}-\eqref{eq:d_opt_det_ET}, LCP can achieve optimality in some cases, \emph{e.g.}, in the presence of deterministic arrivals.

LCP is a policy that, except for the $\min$ operators, does not depend upon the energy status. When the distribution has a small standard deviation, then it is expected that LCP provides good results and moreover, in the deterministic case, it degenerates to an optimal policy.

\section{Offline Optimization} \label{sec:offline}

We now focus on offline optimization. One of the key aspects of this approach is that the energy arrival sequence is assumed to be known a priori (a statistical knowledge of the arrival process is not sufficient). Therefore, we restrict the study to the finite horizon case, considering separately the two cases of infinite and finite batteries. In this context, the aim is to find the Optimal Offline Policy $\mu^\star$ (OP-OFF), \emph{i.e.}, the sequence of actions $(\mathbf{P},\mathbf{D})$ that maximize $G_\mu^K$ (Equation~\eqref{eq:G_H_q_d_fin}). In Section~\ref{sec:real_data} we will use OP-OFF as a benchmark for the online ones in the finite horizon case.\footnote{In this case, we simply apply to the finite horizon scenario the optimal online policy for infinite horizon derived in Section~\ref{sec:online}.}

\subsection{OP-OFF - Infinite Batteries} \label{sec:off_inf}

We first set up the offline optimization problem~\eqref{eq:mu_star} by clearly specifying the constraints that have to be satisfied and the optimization variables, in the case where the battery sizes are infinite. A formulation for the case with finite batteries will be given in the next subsection. 

In this case, the optimization problem in \eqref{eq:mu_star} can be explicitly written as follows (we start with empty batteries):~
\begin{subequations}
\begin{alignat}{3}
  & \min_{\mu = (\mathbf{P},\mathbf{D})} \sum_{k=1}^K -g(P_k) && &&\label{eq:max_g_Q_inf} \\
  & q^{\rm tx}(P_k) \leq E_k^{\rm tx}, \ && k = 1,\ldots,K, &&\label{eq:Q_tx_k}\\
  & q^{\rm rc}(P_k) + D_k \leq E_k^{\rm rc}, \ &&k = 1,\ldots,K, &&\label{eq:Q_D_rc_k} \\
  & P_k \geq 0, \qquad D_k \geq 0, \ && k = 1,\ldots,K, &&\label{eq:Q_D_0} \\
  & E_{k+1}^{\rm tx} = E_k^{\rm tx}-q^{\rm tx}(P_k)\ +\ &&\beta D_k + B_k^{\rm tx}, \quad && k = 1,\ldots,K-1, \label{eq:E_k_tx_inf}         \\
  & E_{k+1}^{\rm rc} = E_k^{\rm rc}-q^{\rm rc}(P_k)\ -\ &&D_k + B_k^{\rm rc}, \ && k = 1,\ldots,K-1, \label{eq:E_k_rc_inf} \\
  & E_1^{\rm tx} = E_1^{\rm rc} = 0. && &&
\end{alignat} \label{eq:off_prob_inf}
\end{subequations}

\noindent Note that the battery evolutions include neither $\min$ operations (because the batteries are infinite) nor $\max$ operations (thanks to~\eqref{eq:Q_tx_k} and~\eqref{eq:Q_D_rc_k}). We recall that the energy harvested in slot $k$ can be exploited only in a later slot and similarly for the energy transferred with ET ($\beta D_k$).

\begin{lemma}
$S \triangleq \{\mu = (\mathbf{P},\mathbf{D}) : \eqref{eq:Q_tx_k}-\eqref{eq:Q_D_0} \mbox{ are satisfied}    \}$ is a convex set.
\begin{proof}
$S$ is a convex set if $q^{\rm tx}(P_k) - E_k^{\rm tx}$, $q^{\rm rc}(P_k) + D_k - E_k^{\rm rc}$, $-P_k$ and $-D_k$ are concave function of $(P_k,D_k)$ for every $k = 1,\ldots,K$. These conditions are satisfied because $q^{\rm i}(P_k)$ are defined as concave functions and the other constraints are linear.
\end{proof}
\end{lemma}

Since the reward function is convex (sum of convex functions) and $S$ is a convex set, \eqref{eq:off_prob_inf} is a convex problem and can be solved using standard optimization techniques.

\subsection{OP-OFF - Finite Batteries}

When the battery sizes are finite, the optimization problem is the same of Equations~\eqref{eq:max_g_Q_inf}-\eqref{eq:Q_D_0}, with the
battery update formulas~\eqref{eq:E_k_tx_inf}-\eqref{eq:E_k_rc_inf} replaced by~
\begin{alignat}{2}
E_{k+1}^{\rm rc} & = \min\{E_k^{\rm rc}-q^{\rm rc}(P_k)- D_k +B_k^{\rm rc},\ && E_{\rm max}^{\rm rc}\}, \\
E_{k+1}^{\rm tx} & = \min \{E_k^{\rm tx}-q^{\rm tx}(P_k)  + \beta D_k +B_k^{\rm tx},\ && E_{\rm max}^{\rm tx} \}.
\end{alignat}

The problem can be formulated in a \emph{standard form} (convex function to minimize plus inequality and equality constraints) by adding an inequality constraint for every possible condition imposed by the $\min$ operations. For example, for the receiver, the first four inequalities that have to be satisfied are ($Q^{\rm i}_k \triangleq q^{\rm i}(P_k)$)~
\begin{subequations}
\begin{align}
Q^{\rm rc}_1 + D_1 \leq & \ 0, \quad 
Q^{\rm rc}_2 + D_2 \leq 
\begin{cases}
E_{\rm max}^{\rm rc}, \\
B_1^{\rm rc}-Q^{\rm rc}_1-D_1,
\end{cases} \\
Q^{\rm rc}_3 + D_3 \leq &
\begin{cases}
E_{\rm max}^{\rm rc}, \\
E_{\rm max}^{\rm rc} + B_2^{\rm rc}-Q^{\rm rc}_2-D_2, \\
B_1^{\rm rc}-Q^{\rm rc}_1-D_1  + B_2^{\rm rc}-Q^{\rm rc}_2-D_2,
\end{cases} \\
Q^{\rm rc}_4 + D_4 \leq &
\begin{cases}
E_{\rm max}^{\rm rc}, \\
E_{\rm max}^{\rm rc} + B_3^{\rm rc}-Q^{\rm rc}_3-D_3, \\
E_{\rm max}^{\rm rc} + B_2^{\rm rc}-Q^{\rm rc}_2-D_2  + B_3^{\rm rc}-Q^{\rm rc}_3-D_3, \\
B_1^{\rm rc}-Q^{\rm rc}_1-D_1 + B_2^{\rm rc}-Q^{\rm rc}_2-D_2 \\
\qquad \qquad \qquad \qquad + B_3^{\rm rc}-Q^{\rm rc}_3-D_3,
\end{cases}
\end{align}
\label{eq:constraints_example}
\end{subequations}

\noindent and similar constraints have to be considered for the transmitter. The total number of
constraints scales as $K^2$.

The general expressions for the transmitter and receiver constraints can be written in compact form as ($i =1,\ldots,k$ and $k = 1,\ldots,K$)~
\begin{align}
\sum_{j = i}^k Q^{\rm tx}_j - \sum_{j = i}^{k-1} \beta D_j & \leq E_{\rm max}^{\rm tx} \chi\{i > 1\} + \sum_{j = i}^{k-1} B_j^{\rm tx}, \\
\sum_{j = i}^k (Q^{\rm rc}_j + D_j) & \leq E_{\rm max}^{\rm rc} \chi\{i > 1\} + \sum_{j = i}^{k-1} B_j^{\rm rc}, \label{eq:constraints_RC}
\end{align}

\noindent where $\chi\{\cdot\}$ is the indicator function. The four cases in~\eqref{eq:constraints_example} can be obtained from~\eqref{eq:constraints_RC} for $k=1, 2, 3, 4$ (note that there are $k$ constraints in each case, obtained for $i=1,\ldots,k$).

For example, when $i = 1$ or $i = k$, the last and the first lines of~\eqref{eq:constraints_example} are obtained, respectively.

In practice, techniques such as the interior-point algorithm or the SQP algorithm can be used to find the optimal solution. However, if the time horizon is large, the computational time can be long. Moreover, to run the algorithms $B_k^{\rm tx}$ and $B_k^{\rm rc}$ must be known in advance. Thus, even if the offline optimization gives the policy with the highest reward among all, in practice it can rarely be used. On the other hand, finding the optimal offline policy is still useful, as it makes it possible to understand what are the limits of the energy transfer mechanism, and can be used as a benchmark for all other policies.

\section{Numerical Results - Online Optimization}\label{sec:long_term}

In this section we present some numerical results for the online policies. In order to understand their properties, here we consider some analytical examples in the infinite horizon case. In Section~\ref{sec:real_data} we discuss how these policies can be applied to a realistic scenario, with finite horizon and real data.

In addition to studying the optimal policy OP-ON, we present the performance of sub-optimal policies in several settings. We remark that, since we focus on the online case, all energies are expressed in terms of energy quanta.

We consider the long-term maximization of $G_{\eta}$ (Equation~\eqref{eq:G_eta} or, equivalently, \eqref{eq:G_H_q_d}) when the reward function is the transmission rate $g(x) = \ln(1+\Lambda x)$, where $\Lambda$ is a scaling factor. $\eta^\star$ is the optimal policy obtained when ET is used, whereas $\eta_0^\star$ is the optimal policy without ET.

\begin{figure}[t]
    \centering
    
    \includegraphics[trim = 1.5mm 0mm 1.5mm 5.5mm,  clip, width=1\columnwidth]{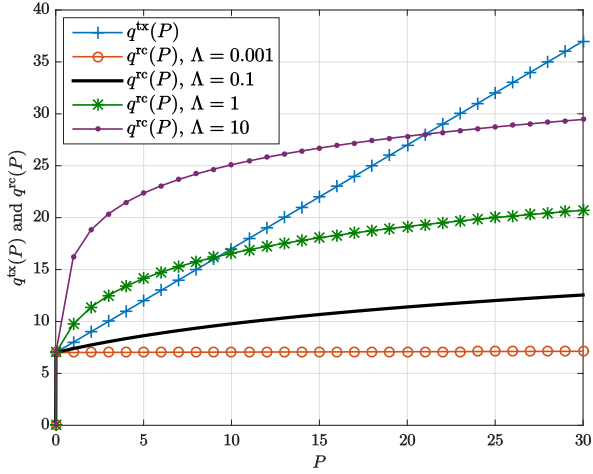}          
    \caption{Energy consumptions $q^{\rm tx}(P)$ and $q^{\rm rc}(P)$ as a function of $P$ for several values of $\Lambda$.}
    \label{fig:q_tx_rc_Lambda}
\end{figure}

The numerical results strongly depend upon the system parameters, and on the structure of $g(\cdot)$ and $q^{\rm i}(\cdot)$. In the following we focus on a particular energy consumption model, but similar considerations can be made in other cases as well. Consider the following energy consumption functions ($\zeta > 0$) expressed in energy quanta~
\begin{align}
q^{\rm rc}(P) = 
\begin{cases}
\frac{\zeta+P_n}{P_n} P, \quad & \mbox{if } P < P_n, \\
\zeta+P_n - \alpha^{\rm rc} \ln(1+\Lambda P_n)& \\
  \ \qquad +\ \alpha^{\rm rc} \ln(1+\Lambda P), \quad & \mbox{if } P \geq P_n
\end{cases} \label{eq:q_zeta_log}
\end{align}

\noindent and $q^{\rm tx}(P)$ is piece-wise linear as in Equation~\eqref{eq:q_ex3} with $P_n = 1/100$. $\zeta$ and $\alpha^{\rm rc}$ are parameters that depend upon the considered technology. Both devices have a fixed energy cost $\zeta$ plus a linear or logarithmic curve.\footnote{We decided to focus on the case $\zeta^{\rm tx} = \zeta^{\rm rc}$ for presentation simplicity, but this assumption is not restrictive.}

If not otherwise specified, we consider $e_{\rm max} \triangleq e_{\rm max}^{\rm tx} = e_{\rm max}^{\rm rc} = 30$, truncated geometric arrivals with parameters $\bar{b}^{\rm tx} =  2$, $b_{\rm max} = 5$ for TX, uniform energy arrivals with parameters $\bar{b}^{\rm rc} = 12.5$, $b_{\rm max}^{\rm rc} =  25$ for RC, $\zeta = 7$, $\alpha^{\rm rc} = 4$, $\Lambda = 0.1$, $\beta = 0.15$, a unit slot length and $\rho_{\rm max} = e_{\rm max}$ (in a slot, potentially, all the stored energy can be consumed).

In \figurename~\ref{fig:q_tx_rc_Lambda}, the bold curve represents the energy consumption $q^{\rm i}(\cdot)$ considered in this example. Note that in the online optimization we consider $q_d^{\rm i}(\cdot) = \left\lceil q^{\rm i}(\cdot) \right\rceil$ as described in Section~\ref{sec:online}.

\begin{figure}[t]
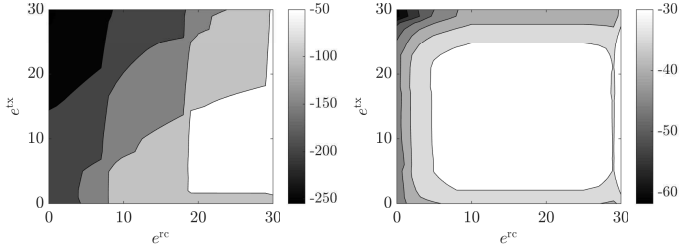

    \centering
    \psfrag{-50}{11}
    \includegraphics[trim = 0mm 0mm 5mm 6mm,  clip, width=0.5\columnwidth]{tx_and_rx_pi_no_ET.eps}~
    \includegraphics[trim = 0mm 0mm 5mm 6mm,  clip, width=0.5\columnwidth]{tx_and_rx_pi_ET.eps}
    \caption{Steady-state probabilities ($10\log_{10}(\cdot)$ scale) without (left) and with (right) ET as a function of the batteries energy status $e^{\rm tx}$, $e^{\rm rc}$.}
    \label{fig:tx_and_rc_pi}
\end{figure}
\begin{figure}[t]
  \centering
  \includegraphics[trim = 1.5mm 0mm 1.5mm 5.5mm,  clip, width=1\columnwidth]{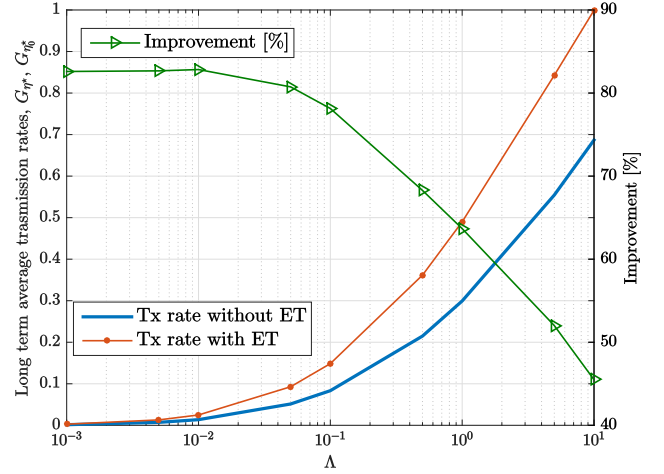}
  \caption{Long-term average transmission rates $G_{\eta_0^\star}$, $G_{\eta^\star}$ (optimal rewards) and corresponding improvement as a function of $\Lambda$ when $e_{\rm max}^{\rm tx} = e_{\rm max}^{\rm rc} = 30$ and $\zeta = 7$.}
  \label{fig:main_Lambda}
\end{figure}

We define the following functions exploiting the technique introduced in Example~\ref{ex:ex3}~
\begin{align}
\begin{split}
\Psi^{\rm tx}(P) &=
\begin{cases}
\frac{1}{m}\ln(1+\Lambda P), \quad &\mbox{if } P < \bar{x}-\zeta, \\
\zeta + P, \quad &\mbox{otherwise},
\end{cases}\\
\Psi^{\rm rc}(P) &= \frac{q^{\rm rc}(\rho_{\rm max})}{\rho_{\rm max}}P,
\end{split}
\end{align}

\noindent with $\rho_{\rm max} = e_{\rm max}-\zeta$, $\bar{x} = 20.99$ and $m = 0.0417$. It can be verified that these functions satisfy the hypotheses of Theorem~\ref{thm:upper_bound_noET} and the upper bounds are $G_{\rm u.b.}^{\rm noET} = 0.0834$ and $G_{\rm u.b.}^{\rm ET} = 0.1561$.

In \figurename~\ref{fig:tx_and_rc_pi} we show the steady-state distribution of the system state when using the optimal policies with and without ET. As expected, when energy transfer is not used, the energy levels are highly unbalanced and the receiver is almost always in overflow. With energy transfer, instead, the overflow probability becomes lower. In this case, even in the presence of a relatively low efficiency, $\beta$ ($85\%$ of the energy sent is wasted), energy transfer provides a reward improvement of $78\%$, see \figurename~\ref{fig:main_Lambda}. Note that the improvement is due to the fact that RC can send part of its energy to TX and this is particularly effective when RC receives more energy and/or consumes less energy than TX. A comparison with the upper bounds shows that $G_{\eta_0^\star} > 0.99 G_{\rm u.b.}^{\rm noET}$ and $G_{\eta^\star} > 0.95 G_{\rm u.b.}^{\rm ET}$. The reward without ET and its upper bound are very close (this happens because the batteries are large). Instead, with ET the distance from the upper bound is wider because the function $\Psi^{\rm rc}(\cdot)$ is distant from $q^{\rm rc}(\cdot)$ and the batteries are not sufficiently large.
When $\Lambda \in \{0.001,1,10\}$ the improvements provided by the use of ET become $\{83,64,45\} \%$, thus the performance is significantly increased in a wide range of values of $\Lambda$. This can be observed in \figurename~\ref{fig:main_Lambda}, where we plot the rewards with and without ET, along with the corresponding improvement, defined as $\left(G_{\eta^\star} - G_{\eta_0^\star}\right)/G_{\eta_0^\star}$. ET works better in the low SNR regime because $g(\cdot)$ tends to be linear, thus smart energy transmission techniques (\emph{e.g.}, delay a transmission in order to transmit with more power) do not improve the reward significantly.

\begin{figure}[t]
  \centering
  \includegraphics[trim = 1.5mm 0mm 1.5mm 5.5mm,  clip, width=1\columnwidth]{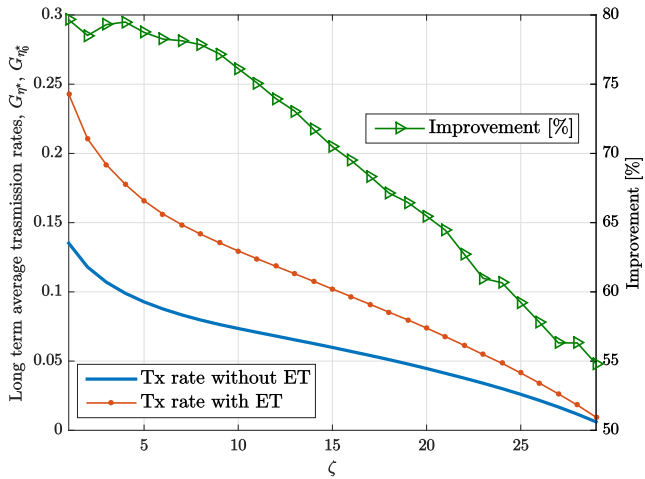}
  \caption{Long-term average transmission rates $G_{\eta_0^\star}$, $G_{\eta^\star}$ (optimal rewards) and corresponding improvement as a function of $\zeta$ when $e_{\rm max}^{\rm tx} = e_{\rm max}^{\rm rc} = 30$ and $\Lambda = 0.1$.}
  \label{fig:main_zeta}
\end{figure}

\figurename~\ref{fig:main_zeta} shows how the two rewards (with and without energy transfer) change as a function of $\zeta$. When $\zeta$ is very high, in both cases the value of the reward is very small in absolute terms (see \figurename~\ref{fig:main_zeta}), but the use of energy transfer may provide a significant reward improvement in relative terms as pointed out by the improvement curve ($G_{\eta^\star} > 1.5 G_{\eta_0^\star}$). Thus, it is better to use Energy Transfer even when $\zeta$ is high. Even if we present our results for $\zeta^{\rm tx} = \zeta^{\rm rc}$, similar results can be found in the general case. In particular, if either energy consumption $\zeta^{\rm i}$ decreases, then the reward improvement and the reward itself increase (similarly to \figurename~\ref{fig:main_zeta}) and vice-versa.

\begin{figure}[t]
  \centering
  \includegraphics[trim = 1.5mm 0mm 1.5mm 5.5mm,  clip, width=1\columnwidth]{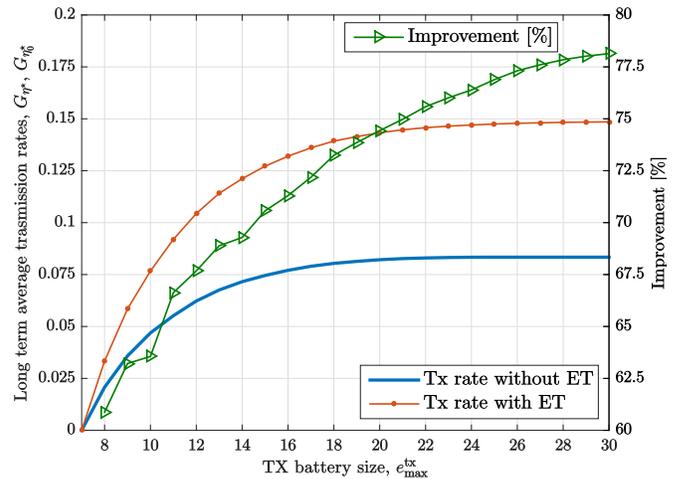}
  \caption{Long-term average transmission rates $G_{\eta_0^\star}$, $G_{\eta^\star}$ (optimal rewards) and corresponding improvement as a function of $e_{\rm max}^{\rm tx}$ when $e_{\rm max}^{\rm rc} = 30$, $\zeta = 7$ and $\Lambda = 0.1$.}
  \label{fig:plot_G_e_max}
\end{figure}

Also, in \figurename~\ref{fig:plot_G_e_max} we plot the reward when $e_{\rm max}^{\rm rc} = 30$ is fixed and $e_{\rm max}^{\rm tx}$ changes (a similar curve can be obtained switching $e_{\rm max}^{\rm tx}$ and $e_{\rm max}^{\rm rc}$). The ET improvement increases with the battery size. The abscissa values start from $7$ since, for $e_{\rm max}^{\rm tx} \leq 7$, the reward is zero because of the circuitry costs.

As an additional interesting example, consider the case $\zeta = 0$, where $q^{\rm tx}(\cdot) = \Psi^{\rm tx}(\cdot)$ and $q^{\rm rc}(\cdot) = \Psi^{\rm rc}(\cdot)$. The energy consumption functions are~
\begin{align}
q^{\rm tx}(P) = P, \qquad q^{\rm rc}(P) = 4 \ln(1+\Lambda P). \label{eq:q_tx_rc_zeta_0}
\end{align}

\noindent In this case $\Psi^{\rm i}(\cdot) = q^{\rm i}(\cdot)$. The distances from $G_{\rm u.b.}^{\rm noET}$ and $G_{\rm u.b.}^{\rm ET}$ are $0.25 \%$ and $3.3 \%$, respectively. For larger batteries the upper bound gaps are even smaller. We also computed the rewards of policies GP, BP and LCP and we found $G_{\rm GP} = 0.88 G_{\eta^\star}$, $G_{\rm BP} = 0.88 G_{\eta^\star}$, $G_{\rm LCP} = 0.82 G_{\eta^\star}$, \emph{i.e.}, in this particular case, the simpler policies provide almost as good a performance as OP, while being significantly faster to compute.

\section{Real Data Analysis}\label{sec:real_data}

\begin{figure}[t]
  \centering
  \includegraphics[trim = 1.5mm 0mm 1.5mm 5.5mm,  clip, width=1\columnwidth]{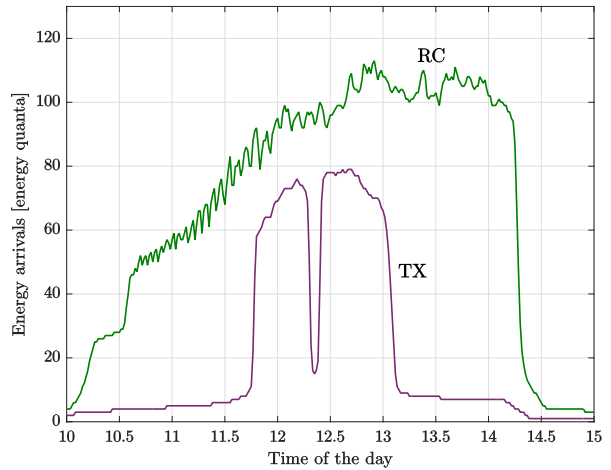}
  \caption{Indoor light energy arrivals as a function of the time of the day.}
  \label{fig:indoor_energy_arrivals}
\end{figure}

\begin{figure}[t]
  \centering
  \includegraphics[trim = 1.5mm 0mm 1.5mm 5.5mm,  clip, width=1\columnwidth]{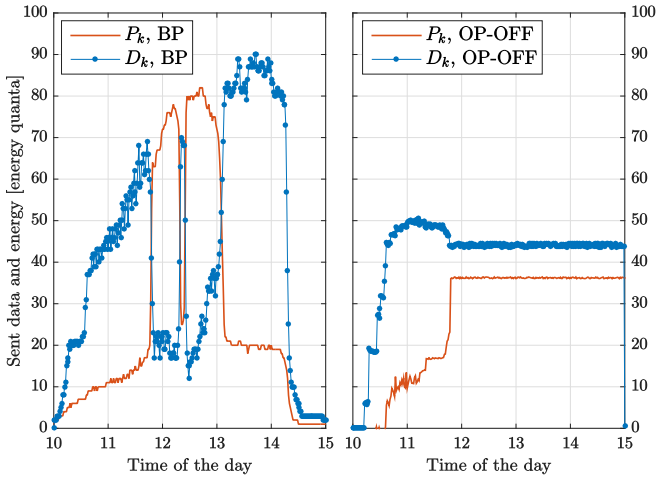}~
  \caption{Policies BP (left) and OP-OFF (right) as a function of the time of the day.} 
  \label{fig:indoor_data_and_energy}
\end{figure}

\begin{figure}[t]
  \centering
  \includegraphics[trim = 1.5mm 0mm 1.5mm 5.5mm,  clip, width=1\columnwidth]{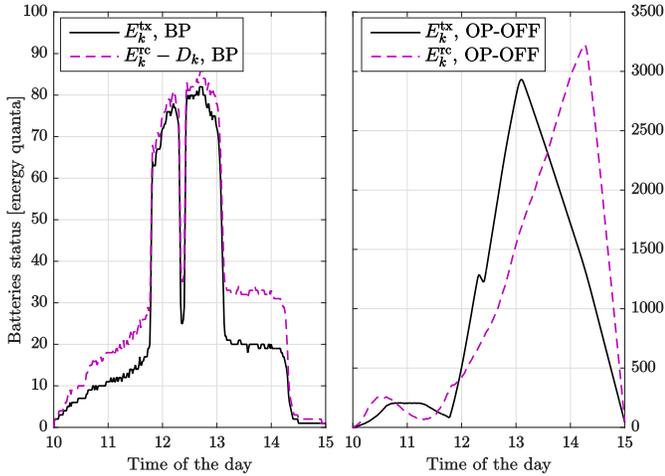}~
  \caption{Battery energy status of BP (left) and OP-OFF (right) as a function of the time of the day.} 
  \label{fig:indoor_batteries_status}
\end{figure}

In this section we want to apply the policies found so far to some realistic examples. Since in reality only a finite sequence of energy arrivals can be available, we focus on the optimization of $G_\mu^K$ (Equation~\eqref{eq:G_H_q_d_fin}). If we assume that the energy arrivals are known a priori, the offline optimal policy (OP-OFF) provides the best reward among all. Instead, to compute the online policies, only the statistics of the energy arrivals is required. In this section, in addition to discussing the benefits of ET, we compare the offline and online approaches. As in Section~\ref{sec:offline}, we consider separately the cases of infinite and finite batteries.

\subsection{Infinite Batteries}
Consider a scenario with two devices in two different rooms of a building, where energy harvesting is based on indoor light.

At enhants.ee.columbia.edu, a collection of light energy data traces is available.\footnote{These data were discussed in~\cite{Gorlatova2013} by Gorlatova \emph{et al.}.} The authors took measurements of the irradiance in different indoor rooms during an extended period of time. We use part of this data in our performance evaluation.

We assume that TX is located on a bookshelf in an office (Setup A) and the receiver in another office (Setup B). The receiver, generally, harvests more energy than the transmitter because it gets more sunshine. We show in \figurename~\ref{fig:indoor_energy_arrivals} the irradiance arrivals for the two devices (measured on $09$ January $2010$). It can be seen that, in this case, RC receives significantly more energy than the transmitter, therefore it may be interesting to use energy transfer to try to balance the system. In this setup, the harvested power is at most $113\ \mu\mbox{W/cm}^2$, \emph{i.e.}, very low. In an indoor environment, an ultra low power sensor network should be deployed, otherwise the energy costs would be too high to be sustained by the renewable energy source. Therefore, we assume that the transmitter can choose its transmit power to be even lower than $1 \ \mbox{mW}$. In this case, it can be verified that the effects of a finite battery can be neglected (even if a very small battery is used, \emph{e.g.}, $0.16\ \mbox{J}$~\cite{Gorlatova2013}), thus in this section we can consider infinite batteries with no loss of generality.

Time is divided in slots of $60\ \mbox{s}$ each, and in every slot a new $(P_k,D_k)$ is chosen. The maximum energy that can arrive in $60\ \mbox{s}$ is $60\ \mbox{s} \times 113\ \mu\mbox{W/cm}^2 \times S \ \mbox{cm}^2$ where $S$ is the solar panel size (assumed equal for the two devices). We compute the reward using $g(x) = \ln(1+\Lambda x)$ in a low SNR regime ($\Lambda = 0.002$). In order to highlight the system behavior, we present the results for $q^{\rm tx}(P) = q^{\rm rc}(P) = P$. The model can be extended, \emph{e.g.}, using the energy consumption model of Equation~\eqref{eq:q_tx_rc_zeta_0}, which would result in an even better improvement because RC would consume less energy.

We use two approaches to apply ET to the system: 1) online low complexity balanced policy (BP), which is very easy to compute and can be used in practice, and 2) offline optimal policy (OP-OFF) (presented in Section~\ref{sec:off_inf}). We selected $1\ \mu\mbox{W/cm}^2$ as the minimum non negligible power that can be harvested. In this case, one energy quantum corresponds to the minimum energy that can arrive in $60\ \mbox{s}$, \emph{i.e.},~
\begin{align}
  1\ \mbox{e.q.} \equiv 60\ \mbox{s} \times 1\ \mu\mbox{W/cm}^2 \times S \ \mbox{cm}^2.
\end{align}

\figurename~\ref{fig:indoor_data_and_energy} shows the sent data and energy (expressed in energy quanta) for BP and OP-OFF. In \figurename~\ref{fig:indoor_batteries_status}, the corresponding energy evolutions are presented.

BP is designed in order to balance the energy of the two devices. Indeed, when the transmitter battery is low, $D_k$ (transfer energy from RC to TX) is high, \emph{i.e.}, ET is better exploited when the difference between the energy arrivals is high. Analytically, it can be verified that in the linear energy consumption case, BP degenerates in the following policy:~
\begin{align}
    d(\mathbf{e}) =\ & \left(\left\lfloor \frac{e^{\rm rc}-e^{\rm tx}}{1+\beta}\right\rfloor\right)^+, \\
    \rho(\mathbf{e}) =\ & \min\{e^{\rm tx},e^{\rm rc}-d(\mathbf{e})\}, \label{eq:BP_rho_lin}
\end{align}

\noindent where $(\cdot)^+ \triangleq \max\{\cdot,0\}$.

On the left side of \figurename~\ref{fig:indoor_batteries_status} we depicted $E_k^{\rm tx}$ and $E_k^{\rm rc}-D_k$ in order to compare the two arguments of Equation~\eqref{eq:BP_rho_lin}. It can be seen that $E_k^{\rm tx}$ is always lower than $E_k^{\rm rc}-D_k$, thus Equation~\eqref{eq:BP_rho_lin} becomes $\rho(\mathbf{e}) = e^{\rm tx}$ (indeed the curves of $P_k$ and $E_k^{\rm tx}$ are the same), \emph{i.e.}, the transmitter battery is emptied in every slot. Moreover, note that $E_{k+1}^{\rm tx} = B_{k}^{\rm tx} + \lfloor\beta D_k\rfloor$, \emph{i.e.}, the status of the transmitter battery is similar to $B_k^{\rm tx}$, but higher (thanks to energy transfer).

Instead, OP-OFF chooses the initial values of $P_k$ and $D_k$ in order to reach a situation where $P_k$ and $D_k$ can be kept constant. This is possible because we consider infinite batteries. The resulting battery trends are represented on the right side of \figurename~\ref{fig:indoor_batteries_status}. Note that $P_k$ and $D_k$ were chosen in order to have zero energy stored in the last plus one slot, \emph{i.e.}, all the available energy is exploited in the finite horizon of $K$ slots. Differently from the previous case, $E_k^{\rm tx}$ is greater than $E_k^{\rm rc}$ in the central region because TX receives a lot of energy and RC transfers its energy to TX.

Note that, if ET is not employed, an upper bound for the performance is given by the minimum between the means of $\{B_k^{\rm tx}\}$ and $\{B_k^{\rm rc}\}$, whereas, if ET is used, the upper bound is given by Equation~\eqref{eq:G_ub_ET}.\footnote{Theorems~\ref{thm:upper_bound_noET} and~\ref{thm:upper_bound_ET} can be reformulated using the temporal means in this case.}

BP gives a reward equal to $0.0512$, whereas $G_{\mu^\star} = 0.0528$ (optimal offline reward with ET) and $G_{\mu_0^\star} = 0.0411$ (optimal offline reward without ET). 
The upper bound with and without ET are $G_{\rm u.b.}^{\rm ET} = 0.0532$ and $G_{\rm u.b.}^{\rm noET} = 0.0414$. Note that $G_{\mu^\star} = 0.99 G_{\rm u.b.}^{\rm ET}$ and $G_{\mu_0^\star} = 0.99 G_{\rm u.b.}^{\rm noET}$, \emph{i.e.}, OP-OFF is very close to but does not achieve the upper bounds even if the batteries are infinite and this is because we consider a finite time horizon. 
The reward improvement due to ET is $28 \%$. 
Note that, even though BP is a sub-optimal policy (much simpler to compute than OP-OFF) and only has a causal knowledge of the energy arrivals, its reward $G_{\rm BP}$ is very close to that of the optimal offline policy, $G_{\mu^\star}$.

\subsection{Finite Battery Effects}

\begin{figure}[t]
  \centering
  \includegraphics[trim = 1.5mm 0mm 1.5mm 5.5mm,  clip, width=1\columnwidth]{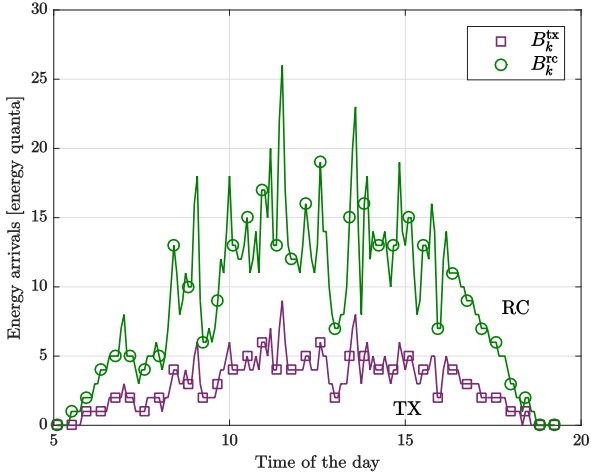}
  \caption{Solar energy arrivals as a function of the time of the day.}
  \label{fig:solar}
\end{figure}

In the previous section we assumed infinite batteries, which is legitimate in the indoor environment we considered. However, when the solar panel is powered with direct sunlight, it is likely that an inappropriate use of the energy may lead to battery overflow. At~\cite{NREL}, a collection of solar light measurements in several locations over the past years is available and in \figurename~\ref{fig:solar} we show the irradiance measured in Elizabeth City on 20 July 2014. The continuous lines represent all the measured data. We performed a sampling and considered only the points depicted with squares and circles. This is in order to perform the offline optimization in a reasonable computational time (we recall that with finite batteries the number of constraints grows quadratically with the number of samples). We considered the same energy arrival profile for both transmitter and receiver, but we assumed that the transmitter has a solar panel three times smaller than RC (in reality, the two devices could also receive different solar energy because of their position). We scaled the irradiance data in order to apply an MDP approach to solve the problem: the histograms of the two energy arrival profiles were assumed as empirical pdfs of the two arrival processes and we found OP-ON according to the model of Section~\ref{sec:online}. Since this approach is sub-optimal because it assumes i.i.d. energy arrivals, we compared it with OP-OFF, that gives the best possible results.

\begin{figure}[t]
  \centering
  \includegraphics[trim = 1.5mm 0mm 1.5mm 5.5mm,  clip, width=1\columnwidth]{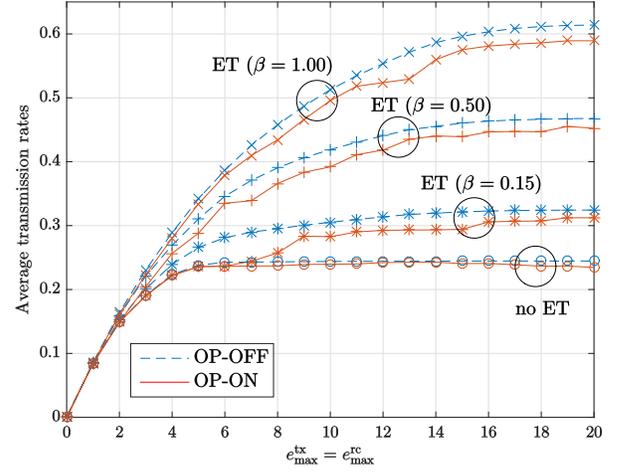}
  \caption{Optimal offline and online rewards as a function of $e_{\rm max}^{\rm tx} = e_{\rm max}^{\rm rc}$ when $\Lambda = 0.1$ and $\beta \in \{0.15,0.50,1.00\}$.}
  \label{fig:solar_G}
\end{figure}

\begin{figure}[t]
  \centering
  \includegraphics[trim = 1.5mm 0mm 1.5mm 5.5mm,  clip, width=1\columnwidth]{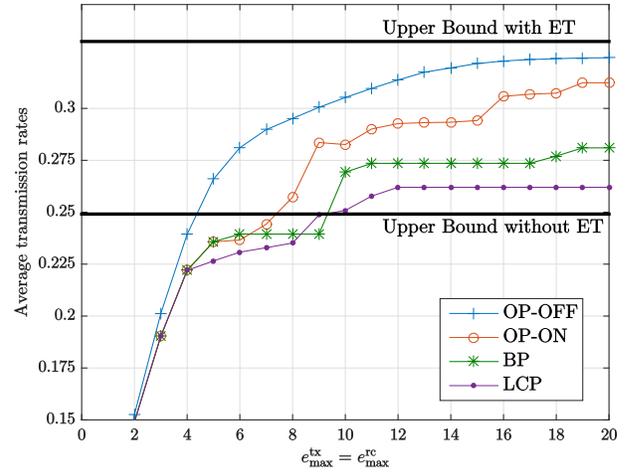}
  \caption{Rewards as a function of $e_{\rm max}^{\rm tx} = e_{\rm max}^{\rm rc}$ when $\Lambda = 0.1$ and $\beta = 0.15$ for several policies.}
  \label{fig:G_policies}
\end{figure}

\figurename~\ref{fig:solar_G} shows the (simulated) rewards with and without ET as a function of the battery sizes. We considered the model of Equation~\eqref{eq:q_tx_rc_zeta_0} with $\Lambda = 0.1$, $e_{\rm max} = e_{\rm max}^{\rm tx} = e_{\rm max}^{\rm rc}$ and $\beta \in \{0.15,0.50,1.00\}$. When $e_{\rm max}$ is low, even when $\beta = 1$, ET does not improve the system reward. This is because the energy harvesting mechanism manages to fill up both batteries almost all the time, thus it is not necessary to exchange energy. Instead, when the size of the batteries grows, ET may significantly improve the reward: when $e_{\rm max} = 5$ the ratio $G_{\mu^\star}/G_{\mu_0^\star}$ for $\beta \in \{0.15,0.50,1.00\}$ is $\{1.12,1.30,1.44\}$, and becomes $\{1.33,1.91,2.51\}$ when $e_{\rm max} = 20$. 

Beyond a certain value of $e_{\rm max}$, the rewards can be observed to saturate, thus it is not necessary to use very large batteries to achieve high rewards. This is because the effects of outage and overflow become negligible. Note that, because of the transmitter energy arrivals, without ET the system reward saturates very soon, whereas, with energy transfer, the saturation value is only reached for higher $e_{\rm max}$.  Note that for $\beta = 0.15$ and $e_{\rm max} \leq 7$, $G_{\eta^\star}$ is low and this is due to the discretization ($\lfloor \beta e_{\rm max} \rfloor = 0$ for $e_{\rm max} < 7$).

In this example OP-ON and OP-OFF are very close, which makes online policies very good candidates for application in real scenarios, because they are easier to implement while being almost optimal.

Finally, in \figurename~\ref{fig:G_policies} we plot OP-OFF, OP-ON, the sub-optimal online policies BP and LCP and the upper bounds. Note that with the online policies OP-ON may be lower than BP (at $e_{\rm max} = 6$ for example). This is because OP-ON is optimal in the \emph{long-run}, thus in a particular realization it may turn out to be sub-optimal. OP-OFF increases with $e_{\rm max}$ and almost reaches the upper bound (which is not achieved because the simulation time is finite). The Balanced Policy is generally better than the Low Complexity Policy, because BP operates with the energy levels (see Equation~\eqref{eq:BP}), whereas LCP operates with the average energy arrival statistics (see Equation~\eqref{eq:LCP}).

\section{Conclusions}\label{sec:conclusions}
In this paper we jointly analyzed two mechanisms, namely Ambient Energy Harvesting and Wireless Energy Transfer, that can be used to improve the network performance. We studied a scenario composed of two Energy Harvesting Devices, a transmitter and its receiver, that can exchange energy through an Energy Transfer interface. We considered two generic energy consumption functions and found performance upper bounds with and without ET, showing that, under some assumptions, they are achievable. Then we studied the online and offline optimization problems. In the first case we modeled the system with an MDP, studying numerically the optimal online policy and introducing some low complexity policies. For the offline optimization we set up the optimization problem and showed that it is convex. In our numerical evaluations we derived the optimal transmission policies, showing that ET can significantly improve the system performance and discussing how the system behaves as a function of the system parameters. For example, we noticed that the reward improvement increases with the battery sizes and remains high even for large values of the circuitry cost. Also, we analyzed two realistic examples of indoor and outdoor light radiation, showing the effects of finite batteries on the transmission strategies. Possible extensions of our work are the exploitation of the predictability and correlation of the transmitter and receiver energy sources, and consideration of battery imperfections.
\appendices

\section{Proof of Theorem~\ref{thm:upper_bound_noET}} \label{app:upper_bound_noET}

The energy harvesting mechanism imposes~
\begin{align*}
\limsup_{K \rightarrow \infty} \frac{1}{K}\sum_{k = 1}^K q^{\rm i}(P_k) \leq \bar{b}^{\rm i}.
\end{align*}

Using the definitions~\eqref{eq:G_H_q_d_fin}-\eqref{eq:G_H_q_d} and the hypotheses, we have~
\begin{align*}
G_\mu =& \limsup_{K \rightarrow \infty} \frac{1}{K}\sum_{k = 1}^K g(P_k) \\
=& \limsup_{K \rightarrow \infty} \frac{1}{K}\sum_{k = 1}^K g(\Psi^{{\rm i}^{-1}}(\Psi^{\rm i}(P_k))) \\
\leq& \limsup_{K \rightarrow \infty} g\left(\Psi^{{\rm i}^{-1}} \left(\frac{1}{K}\sum_{k = 1}^K \Psi^{\rm i}(P_k) \right)\right) \\
\leq& g\left(\Psi^{{\rm i}^{-1}} \left( \limsup_{K \rightarrow \infty} \frac{1}{K}\sum_{k = 1}^K q^{\rm i}(P_k) \right)\right)
\leq g(\Psi^{{\rm i}^{-1}} (\bar{b}^{\rm i} )).
\end{align*}

The relation holds for both TX and RC, thus, since we deal with increasing functions, \eqref{eq:G_ub_noET} is obtained.

For the last point of the theorem we introduce the following proposition.
\begin{propos}
    If $\Psi^{\rm i}(P)$ does not exist, then the battery of device ${\rm i}$ is infinite.
    \begin{proof}
        We will equivalently show that if the battery size is finite, then $\Psi^{\rm i}(\cdot)$ always exists. Since the battery is finite, the transmission power is bounded by $\rho_{\rm max} < \infty$. The function $\Psi^i(\cdot)$ can be chosen as a linear function $\Psi^{\rm i}(P) = m P$ where $m$ is a slope such that $m P \leq q^{\rm i}(P)$. Thus, since $\Psi^{\rm i}(\cdot)$ is linear, also its inverse is linear. In this case $g(\Psi^{{\rm i}^{-1}}(\cdot))$ is concave because $g(\cdot)$ is concave, therefore $\Psi^{\rm i}(\cdot)$ can be correctly defined and therefore always exists.
    \end{proof}
\end{propos}

Now, assume that both $\Psi^{\rm tx}(P)$ and $\Psi^{\rm rc}(P)$ do not exist. This implies that the battery sizes are infinite and in this case $g(q^{{\rm i}^{-1}}(P))$ for large $P$ increases faster than $P$ (otherwise $\Psi^{\rm i}(P)$ can be found). To show that the reward tends to infinity, consider the following policy over a time horizon of $K$ slots:~
\begin{align*}
P_1 &= P_2 = \ldots = P_{K-1} = 0, \qquad P_K = q^{{\rm i}^{-1}}\left(\sum_{k= 1}^{K-1} B_k^{\rm i} \right).
\end{align*}

The corresponding reward is~
\begin{align*}
G_\mu^K = \frac{1}{K} g\left( q^{{\rm i}^{-1}}\left(\sum_{k= 1}^{K-1} B_k^{\rm i} \right) \right)
\end{align*}

\noindent and $\lim_{K \rightarrow \infty} G_\mu^K = \infty$ because the argument of $q^{{\rm i}^{-1}}(\cdot)$ grows linearly in $K$.

\bibliography{EHD}{}
\bibliographystyle{IEEEtran}

\end{document}